\documentclass[12pt]{article}

\usepackage{exscale}
\usepackage{amsmath, amssymb, amsthm}

\newcommand {\half} {{1 \over 2}}

\newcommand {\bra}[1] {\left< #1 \right|}
\newcommand {\ket}[1] {\left| #1 \right>}
\newcommand {\bbra}[1] {\left<\left< #1 \right.\right|}
\newcommand {\kket}[1] {\left| \left.#1 \right>\right>}
\newcommand {\eexpect}[1] {\left<\left< #1 \right>\right>}

\newcommand {\braket}[2] {\left< #1 \mskip1mu\vrule\mskip1mu #2 \right>}
\newcommand {\bbraket}[2] {\left<\left< #1 \mskip1mu\vrule\mskip1mu #2 \right>\right>}

\theoremstyle{remark}

\theoremstyle{definition}

\begin{document}

\title{
        Non-perturbative  Quantization 
        of Phantom and Ghost Theories:  Relating Definite and
        Indefinite Representations
      }
    \author{Andr\'e van Tonder${}^{(1)}$ and Miquel Dorca${}^{(1,2)}$
            \\ \\
            ${}^{(1)}$ Department of Physics, Brown University \\
            Box 1843, Providence, RI 02912, USA \\
            \\
	    ${}^{(2)}$ Department of Physics, University of Rhode Island \\
            2 Lippitt Road, Kingston, RI  02881-0817, USA
            }
    \date{Oct 27, 2006}

    \maketitle

    \begin{abstract}
        \noindent
We investigate the non-perturbative quantization of phantom and ghost 
degrees of freedom by relating their representations in definite
and indefinite inner product spaces.  For a large class of potentials,
we argue that the same physical information can be extracted from 
either representation.  We provide a definition of
the path integral for these theories, even in cases where the integrand
may be exponentially unbounded, thereby removing some previous 
obstacles to their non-perturbative study.  We 
apply our results to the study of ghost fields of Pauli-Villars
and Lee-Wick type, and we show in the context of a toy model how 
to derive,  from an exact non-perturbative path integral calculation,  
previously ad hoc prescriptions for Feynman diagram contour integrals 
in the presence  of complex energies.   We point out that the pole
prescriptions obtained in ghost theories are opposite to what would 
have been expected if one had added conventional $i\epsilon$ 
convergence factors in the path integral.

        %\noindent
        %\rule{4.6in}{1pt}
        %\newline
        %{Keywords:}
          
        %\newline
        %arXiv.org eprint: hep-th/
        %\newline
        %{Brown preprint:} BROWN-HET-
        %\newline
        %{PACS:}
        %  
    \end{abstract}

\section{Introduction}

In this article we discuss the non-perturbative quantization 
of bosonic phantom degrees of freedom.  
A phantom is defined as a degree of freedom for
which the sign of the kinetic term is opposite to that of an ordinary
bosonic degree of freedom.  
Phantoms may often be quantized using either an ordinary 
Hilbert space representation or an indefinite inner product representation.
In the latter case, the phantom is usually called a ghost.

It is often physically required that we represent phantom degrees of freedom 
on an indefinite inner product space. 
Examples include the timelike components of 
 gauge fields,
 Faddeev-Popov ghosts, Pauli-Villars regulator ghosts,
and the regulator ghost fields postulated in Lee-Wick \cite{leewick, lee}
ultraviolet completions of field theories.   

To describe such degrees of freedom non-perturbatively,
Boulware and Gross \cite{BG} constructed
a functional integral from an indefinite inner product
representation of the canonical commutation relations.
The resulting functional integral had two main problems, both of which
 we address
in this article.  

The first problem occurs if the 
potential has odd terms.  The integrand then 
 becomes exponentially 
unbounded and 
the functional integral naively diverges, seemingly casting the 
non-perturbative existence of the theory in doubt \cite{BG}.  
This problem
already arises in the presence of source terms that
are linear in the fields \cite{sakoda}.   
We show in this article that such linear -- and in 
certain cases even quadratic -- exponentially unbounded integrands 
can be treated non-perturbatively by defining path 
integrals in the distributional sense on appropriate 
test function spaces of a type studied
by Gel'fand and Shilov in their theory of generalized functions \cite{gelfand1, gelfand2}.      

The second problem occurs when quantizing relativistic fields
using the indefinite metric representation, since the 
region of integration
may not be manifestly covariant.  This happens 
for the electromagnetic field, as discussed by Arisue et al. 
\cite{arisue},
and for the Dirac boson discussed in the present article.  
In particular, the region of integration for each ghost degree of 
freedom is effectively the  imaginary axis.    
In general, there is no coordinate-independent 
 way of distinguishing 
ordinary and ghost degrees of freedom, so that the functional integral 
is not manifestly covariant, though it may still be implicitly covariant
\cite{arisue}.  The main result of the present article overcomes this
problem by showing that information appropriate to the indefinite inner
product representation can in fact be extracted in the usual way
from the usual 
Hilbert space representation, where covariance of the integrand is 
manifest.  

For gauge fields, the existence of a well-behaved analytic continuation 
to Euclidean time makes these  issues less urgent.  However,
for  fields such as the Dirac boson, 
which  are of particular interest as Pauli-Villars and 
Lee-Wick ghosts, the 
Euclidean functional integrand is exponentially unbounded.  
For these fields,
the real-time formalism works fine, as we illustrate. 
Astonishingly, for fields that appear at most 
quadratically in the action, even the exponentially unbounded Euclidean
functional integral can be calculated in an appropriate 
distributional sense, as we discuss in section \ref{general}. 

It is in fact 
common practice to derive results based on heuristic functional integrals,
in theories containing indefinite-metric gauge field components
or ghosts, by using the classical covariant action.  
One may justifiably ask why not.  In the real-time 
formalism, the integrand is $\exp (iS)$.  As long as $S$ is 
real, negative terms in $S$ are not necessarily worse than positive 
terms in $S$, and the 
oscillatory integral may be  equally well defined
in either case.  

In this article we justify the use of the ordinary covariant
action and integration region
 for potentials sufficiently general
to cover the 
applications 
mentioned in the introductory paragraphs.  
In particular,  we show that 
the indefinite-metric ground state expectation values can be 
obtained from the definite-metric transition function 
using the usual $T\to -i\infty$ prescription.

Because they lacked a non-perturbative definition of the indefinite
inner product
theories corresponding to the unbounded path integrands that we are now 
able to 
treat, 
various authors \cite{leewick, lee, cutkosky} studied the consistency of 
ad hoc prescriptions for defining
the propagators, and more complex diagrams, order by order 
in perturbation theory.  
Lee and Wick \cite{leewick, lee} introduced one such prescription,
 requiring the contour 
integrations defining various diagrams to be continuously deformed 
to avoid the movement of poles in the complex plane
as one changes the parameters of the theory. 
 Cutkosky, Landshoff, Olive and Polkinghorne \cite{
cutkosky}
attempted to generalize this prescription to coalescing 
singularities not covered by Lee and Wick.  

The framework of this article 
allows a non-perturbative functional integral to be calculated
in the distributional sense  
for various kinds of unbounded path integrands, including the ones
studied by Lee and Wick.  In principle, we expect the ad hoc 
prescriptions of the above authors to be either reproduced or corrected
in our non-perturbative framework.  We motivate this statement with 
a couple of simple exact calculations in a toy model, where we succeed 
in deriving Lee-Wick-type pole prescriptions from first principles 
via explicitly non-perturbative calculations.  

\section{Related work and applications}

In addition to the seminal references discussed in the introduction, we
also point out the following related work.   

The study of phantoms has recently been of interest in cosmological 
model building.  They are used, for example, in some
 models of cosmological dark energy.  The literature on the subject is
too large to cite representatively, but see for example
 \cite{caldwell}.  There are 
open questions regarding the consistency and stability 
of these models \cite{cline, trodden}.
It is our hope that the non-perturbative methods introduced in the current
article may throw further light on the subject.  For example, in section
\ref{general} we learn how to non-perturbatively treat unbounded and 
complex potentials, which may be useful in studying, from 
a quantum-mechanical point of view, the stability 
questions discussed in, for example, \cite{kujat} and \cite{dabrowski}.  

In \cite{erdem1, erdem2} Erdem, and in \cite{hooft} 't Hooft and Nobbenhuis discuss a novel
kind of symmetry transformation consisting of a rotation of real
positional coordinates to the imaginary axis, with the aim
of ruling out a cosmological constant.  The rotated 
representation that these authors use for their non-relativistic particle 
toy modle is identical to the one we 
discuss in section \ref{freesect} for the indefinite inner product theory.  
These authors are particularly interested in the relationship between 
the real and imaginary coordinate representations.  Since we study 
this relationship in 
detail in the present article,
it is conceivable that our mathematical framework may have further 
applications in this direction.  

Also related to the cosmological constant problem is the paper
\cite{kaplan}, which introduces phantom fields to cancel the 
ordinary matter contribution to the vacuum energy.  Again, our 
non-perturbative approach to phantom fields may be useful in the 
study of these models.  

In the paper
\cite{antoniadis}, Antoniadis et al.\ promote
 the study of theories with ghosts
in the real-time formalism, as opposed to the Euclidean formalism.  
In the real-time formalism, 
they study the 
issue of $i\epsilon$ prescriptions for
resolving ambiguities in Feynman diagrams.  
However, we disagree with the assumption of 
their approach, which derives pole prescriptions by
introducing $i\epsilon$ terms to provide convergence factors in 
path integrals, and states (we believe incorrectly) that the convergence terms
are necessary and that 
theories without such convergence terms are ambiguous.  As we
discuss in section \ref{polesect}, the pole prescription for ground 
state expectation values in ghost theories is 
in fact  \textit{opposite} to what one would expect from such a 
convergence factor.  Our derivation is unambiguous, and does not 
require a convergence factor, since our path integral is well-defined
without it.  
As a result of their
assumptions, the authors of \cite{antoniadis}
find opposite $i0$ prescriptions for the 
 two-point functions of a particle
and a corresponding ghost, in conflict with our results, 
which agree 
with many authors starting with Pauli and Villars \cite{PV}.  
In the light of these remarks, we believe that the conclusions of
\cite{antoniadis} may have to be revisited or reinterpreted.  In this regard, 
we believe that the expectation values calculated in \cite{antoniadis}
may be valid in a representation based on a vacuum that is not 
a ground state.  It is unclear to us how such a vacuum can be 
defined invariantly.

In section \ref{leewicksect}, we provide further non-perturbative 
calculations demonstrating what happens when 
the poles may travel in the complex plane. 

There is a large body of work in scattering theory that 
includes resonances in the description of quantum systems
via complex-coordinate or complex-momentum methods
that displace either the position or
the momentum representation into the complex plane
\cite{berggren, newton, buchleitner, moiseyev, hagen, ho}.
In these cases, the original and displaced
representations are not equivalent, yet they
are expected to encode the same physical information.  What 
we do in the present article may be seen as a special case of
this method, made rigorous for a specific class of potentials.  We
show that the same information is encoded by the representation
based on the definite inner product, whose configuration space is 
the real line, and the representation based on an indefinite 
inner product whose configuration space is the imaginary axis.  

In scattering theory, so-called Siegert or Gamow states 
may be used to represent resonances.  These are states with 
complex momentum \cite{berggren}, 
which may be given precise mathematical meaning
in the framework of section \ref{general} of the current article,
where such states are defined as distributions on test function
spaces of Gel'fand-Shilov type.  Since we are mainly interested
in field theory applications where interactions are polynomial,
we do not treat sufficiently general
potentials for our results to be directly applicable to many
traditional non-relativistic scattering problems, but perhaps 
the mathematical machinery can be generalized.  

The author of \cite{madrid} describes a treatment of these states
in the context of rigged Hilbert spaces.  His construction should be
very closely related to the one of the current article.  

We also mention the following recent work on alternative
 physical interpretations of theories of Lee-Wick 
type, which differs from the approach of the original authors
\cite{bender1}.

\section{Relating definite and indefinite quantizations: The free particle}
\label{freesect}

We will show that certain quantum systems can be 
quantized in either a definite or an indefinite inner product 
representation.  We will also argue that, under certain conditions,
 the choice is a matter of
convenience, since physical quantities appropriate to one 
representation can be obtained from results calculated in the 
other.  

To introduce the ideas, in this section we start with
the ordinary free particle in one dimension with 
Hamiltonian
$$
H = {p^2\over 2m}.
$$
The transition function is given by 
\begin{align}
  \bra{y}e^{-iHt}\ket{x} &= 
     \int {dk \over 2\pi}\,\bra{y}  e^{-iHt}\ket{k}\braket{k}{x} \nonumber\\
     &=  \int {dk\over 2\pi}\,e^{ik(y-x) -itk^2/2m} \label{fourier0}\\
     &= e^{-i\pi/4}\,{\sqrt {m\over 2 \pi t}}\, e^{im(y -x)^2/2t}, \nonumber
\end{align}
where the last line is calculated for positive $t$.
Here we have summed over a complete generalized basis
\begin{align*}
  &\braket{x}{k} = e^{ikx},\quad p\ket k = k \ket k,\quad\braket{k'}{k} = 2\pi\, \delta(k'-k), && k\in \mathbf R,
\end{align*}
of a Hilbert space with positive definite inner product.  The
energy spectrum is positive, and the oscillatory integral has a rigorous 
interpretation as the Fourier transform of a distribution.

Notice that the form (\ref{fourier0}) of the spectral representation
may be continued
to a function analytic in $t$ on the lower half of the complex plane.

The same Hamiltonian can be quantized in an indefinite inner product 
representation.
We will recast the representation used by  Boulware and Gross \cite{BG}
in a more convenient notation similar to the one used by Arisue et al \cite{arisue}.
This proceeds by representing $q$ and $p$ as Hermitian operators 
in an indefinite inner product space spanned by
generalized bases
\begin{align*}
 & p \kket {ik} = ik \kket{ik}, \quad \bbraket{-ik}{ik'} = 2\pi\, \delta(k'-k), \quad &&k\in \mathbf R, \\
&q \kket {ix} = ix \kket{ix}, \quad \bbraket{-ix}{ix'} = \delta(x'-x), \quad &&x\in \mathbf R,
\end{align*}
where
\begin{align*}
  &\bbraket{ix}{ik} = e^{-ikx}.
\end{align*}
Note that the eigenvalues of the Hermitian\footnote{Hermitian operators
on indefinite inner product spaces are sometimes also
called pseudo-Hermitian.  We will not use this qualifier.} operators 
 $q$ and $p$ occur in complex conjugate pairs,
which is allowed in an indefinite inner product 
space  \cite{bognar}.  The corresponding pairs of generalized 
 eigenstates are dual null
states.  The configuration space
is now the imaginary axis, where wave functions take the values
$$
 \phi(ix) \equiv \bbraket {ix}{\phi}.
$$
Thus, a generic wave function in the position representation
has as its domain the imaginary axis.   There is no assumption 
of analyticity -- we cannot analytically extend a generic wave function
 away from the imaginary axis, so that $\bbraket {z}\phi$ is 
in general undefined
unless $z\in i\mathbf{R}$.  We will, however, have reason below to 
investigate subspaces consisting of test functions that are analytic.  
 
The completeness
relations in terms of these sets of states are
$$
    \int {dx}\, \kket {-ix}\bbra{ix} = 1 = \int {dk\over 2\pi}\, \kket {-ik}\bbra{ik}
$$
corresponding to the inner product
$$
   \bbraket{\psi}{\phi} = \int dx\, \left(\psi(-ix)\right)^* \phi(ix). 
$$
satisfying 
$$
  \bbraket{\psi}{\phi} = \bbraket{\phi}{\psi}^*.
$$
This inner product is indefinite.  The state space is 
 not a Hilbert space but a Kre\u\i n space, which may be constructed 
by completing the space of square-integrable functions on the 
imaginary axis with respect to an auxiliary  $\mathcal{L}^2$ inner product to 
obtain a topologically complete vector space, which is then 
equipped with the above indefinite inner product, discarding the
auxiliary  $\mathcal{L}^2$ 
inner product \cite{bognar}.  We emphasize that all that matters 
about the auxiliary
$\mathcal{L}^2$ inner product used in the intermediate step is 
the resulting topological vector space.  The auxiliary inner product 
 is not unique and has no 
physical meaning.
The construction is analogous to using an arbitrary and non-physical Euclidean 
auxiliary inner product to define the topology of Minkowski space.  

Our goal is to investigate the relationship between 
definite and indefinite quantizations of the same Hamiltonian.  
We will therefore compare the
ordinary Hilbert space transition function
$\bra y e^{-iHt} \ket {x=0}$ to the
quantity $\bbra y e^{-iHt} \kket{x=0}$ calculated in the indefinite 
representation, where a suitable meaning
for $\bbra{y}$, $y\in\mathbf{R}$, will be assigned below.  

At first glance, this might not seem like a reasonable thing to do. 
The two quantities are distributions on
different test spaces.  As distributions, 
they are therefore
incomparable.  
However, as long as there exists a range of $t$ for which the 
distributions are 
representable by continuous functions in $y$, we can compare these functions.

The first observation is that
we need a final state $\bbra{y}$, $y\in\mathbf{R}$, whereas
generic states in the indefinite representation are  
functions $\phi (ix) \equiv\bbraket{ix}{\phi}$  
defined on the imaginary 
axis.  Generalized functions such as 
 $\bbra{y}$ can, however, be defined provided we can find a subspace
of the state space consisting of test functions  that may be 
analytically continued from the imaginary axis 
into the complex plane and are, at the same
time, invariant under time evolution.  
For the free particle, a simple test function space satisfying these 
conditions is the
Gel'fand-Shilov space $S^{1/2}_{1/2}(ix)$ \cite{gelfand1}, to be discussed 
in more detail in section \ref{general}.
This test space consists
of entire functions satisfying the 
growth condition\footnote{The space $S^{1/2}_{1/2}$ defined by Gel'fand and Shilov is based on imposing growth conditions on the real axis and corresponds, in the general notation introduced in the section \ref{general}, 
to $S^{1/2}_{1/2}(x)$.  The space $S^{1/2}_{1/2}(ix)$
used here is simply obtained by imposing the same growth conditions on the 
imaginary axis, exchanging the roles of $x$ and $y$.}
\begin{align}
  |y^k \phi(x + iy)| \le C_k \, \exp \left({-a|y|^{2} + b|x|^{2}}\right), \label{growthab0}
\end{align}
where  $C_k$, $\alpha$, and $b$ are positive real numbers that are
allowed to depend on 
$\phi$.\footnote
{Although the powers of $x$ and $y$ in the growth conditions
are preserved under time evolution, it should be noted that the sign of $b$ 
is not, so that the space $S^{1/2}_{1/2}(ix)$ is not quite 
closed under $e^{-iHt}$.  This makes it
necessary to consider instead 
families of generalized functions on 
subspaces denoted by $S^{1/2, B}_{1/2, A}(ix)$ that are mapped into 
each other under time evolution, where $A$ and $B$ are related to 
$a$ and $b$ in (\ref{growthab0}).  The method is developed rigorously
by Gel'fand and Shilov in \cite{gelfand2}, and can be applied  
as long as the powers of $x$ and $y$ in the growth 
condition are preserved under time evolution.  In the interest of brevity,
we will commit a slight abuse by leaving this complication implicit in the rest of this paper.}    

For arbitrary $z\in \mathbf{C}$, we may now define the generalized function 
$\bbra z$ as
$$
  \bbraket{z}{\phi} \equiv \phi(z), \quad \phi \in S^{1/2}_{1/2}(ix).
$$ 
Acting on an element of $S^{1/2}_{1/2}(ix)$, we have, purely formally,
\begin{align*}
    \bbra{y}e^{-iHt}\kket{\phi}
      &=  \int{dk} \bbra{y}e^{-iHt}\kket{-ik}\bbraket{ik}{\phi} 
 \\
&= \int{dk} \bbra{y}e^{-iHt}\kket{-ik}\int d x \bbraket{ik}{-ix}\bbraket{ix}{\phi}.
\end{align*}
However, neither of the supposed kets 
$\kket{-ik}$ nor $\kket{-i\tilde x}$ is a test function, 
and the above should be read as a shorthand for the rigorously defined 
\begin{align*}
    \bbra{y}e^{-iHt}\kket{\phi}
      &=  \int{dk} \,e^{ik^2t + ky} \bbraket{ik}{\phi} \\
&\equiv \int{dk} \,e^{ik^2t + ky} \int dx \,e^{-ikx}{\phi(ix)}.
\end{align*}
It is here that the utility of $S_{1/2}^{1/2}(ix)$ comes to the fore.
Despite the exponentially growing factor $e^{ky}$, the integral over
$k$ converges, since  $S_{1/2}^{1/2}(ix)$ is closed under the Fourier 
transform \cite{gelfand1}, schematically $\mathcal{F}(S_{1/2}^{1/2}(ix)) = 
S_{1/2}^{1/2}(ik)$, which means that  $\bbraket{ik}{\phi}$ decreases 
as $e^{-bk^2}$ for some positive $b$.  Then
\begin{align}
    \bbra{y}e^{-iHt}\kket{\phi}
      &=  \lim_{\epsilon\to 0}\int{dk} \,e^{ik^2t + ky - \epsilon k^2} \int dx \,e^{-ikx}{\phi(ix)}\nonumber\\
&=  \int dx \left( \lim_{\epsilon\to 0}\int{dk} \,e^{ik^2t + k(y - ix) - \epsilon k^2}\right){\phi(ix)}, \label{unusualfunny}
\end{align}
where the added convergence factor, entirely superfluous in the convergent 
integral on the first line, has allowed us to exchange integrations
in the second line.  

We see from the second line that the generalized function
$\bbra{y}e^{-iHt}$ can be represented by the kernel 
\begin{align}
   \bbra{y}e^{-iHt}\kket{- ix} 
    &\equiv \lim_{\epsilon\to 0}\int{dk} \bbra{y}e^{-iHt}\kket{-ik} e^{-\epsilon k^2} \bbraket{ik}{- i x} \nonumber\\
    &= \lim_{\epsilon\to 0}\int{dk} \, e^{ik^2t+k(y- ix) -\epsilon k^2},
      \label{fourier2} \\
   &= e^{i\pi/ 4} \sqrt {m\over 2\pi t}\, e^{im(y - ix)^2 / 2t},
\end{align}
where again the formally undefined 
inner products on the first line should be taken
as a useful mnemonic for the well-defined expression on the second line,
copied from (\ref{unusualfunny}).  
Again, the last line was calculated for positive $t$.  

We have obtained a spectral representation of the kernel
$ \bbra{y}e^{-iHt}\kket{- ix}$ in the indefinite inner product representation.
The expression (\ref{fourier2}) in fact defines 
 a function analytic in $t$ on the upper half 
plane.   

For $x = 0$, we may now compare this with the definite representation, 
and we find 
\begin{align*}
  \bbra{y}e^{-iHt}\kket{x=0}&=
    i \,\bra{y}e^{-iHt}\ket{x=0}
\end{align*}
for positive $t$.  For negative $t$, one may check that
\begin{align*}
  \bbra{y}e^{-iHt}\kket{x=0}&=
    - i \,\bra{y}e^{-iHt}\ket{x=0}.
\end{align*} 
To summarize, we have obtained the following structure:
\begin{itemize}
\item
The function $i\bra{y}e^{-iHt}\ket{x=0}$, calculated in the definite
   inner product representation, has a positive-energy 
spectral representation that 
  defines a function analytic
  in $t$ on the lower half plane.
\item
The function $\bbra{y}e^{-iHt}\kket{x=0}$, calculated in the indefinite
   inner product representation, 
has a negative-energy spectral representation that 
  defines a function analytic in $t$ on the upper half plane.
\item
On the positive real axis in $t$, these two functions coincide.  
Together, they therefore
define a single function analytic on the entire complex plane 
except for a cut on the negative real axis, where the value of 
$\bbra{y}e^{-iHt}\kket{x=0}$
is obtained by approaching the cut from above, and 
$i\bra{y}e^{-iHt}\ket{x=0}$ is obtained by approaching the cut from below.    
\end{itemize}
We see that, for positive real $t$, the results 
calculated using the two representations coincide.\footnote{Since 
the inner product in the indefinite representation has no 
canonical choice of overall sign, we could have made these two quantities 
coincide for negative $t$, and positioned the cut at positive $t$,
by defining  the inner product with an opposite sign.}  
If we knew beforehand that the two 
spectral representations determined the two halves of a single function 
analytic
in $t$, we could have used either representation to infer the 
result for the other.  In section \ref{general}, we will discuss under
what conditions this can be done.

\section{The phantom free particle}
\label{phantomfree}

We briefly discuss the quantization of the particle
with opposite sign Hamiltonian
$$
   H = - {p^2\over 2m}.
$$ 
Now the positive-definite inner product momentum
eigenstates have negative energy
$$
  E_k = - k^2/2m,
$$
and 
\begin{align}
  \bra{y}e^{-iHt}\ket{x} &= 
     \int {dk \over 2\pi}\,\bra{y}  e^{-iHt}\ket{k}\braket{k}{x} \nonumber\\
     &=  \int {dk\over 2\pi}\,e^{ik(y-x) +itk^2/2m} \label{fourier3} \\
     &= e^{i\pi/4}\,{\sqrt {m\over 2 \pi t}}\, e^{-im(y -x)^2/2t}. \nonumber
\end{align}
Note that, although the energy is unbounded below, the real-time 
representation exists.  The spectral representation (\ref{fourier3}) 
now directly defines an analytic
function in $t$ on the upper half plane.

The same transition function can be written in terms of a positive
energy spectral representation  using the
indefinite inner product space defined in section \ref{freesect}.
The states
$$
  \kket{ik}, \quad k\in \mathbf R,
$$
are now eigenstates of $H$ with positive energy $k^2/2m$.  Similarly to 
section \ref{freesect}, we find the spectral representation
\begin{align}
   \bbra{y}e^{-iHt}\kket{- ix} 
    &\equiv \lim_{\epsilon\to 0}\int{dk} \bbra{y}e^{-iHt}\kket{-ik} e^{-\epsilon k^2} \bbraket{ik}{- i x} \nonumber\\
    &= \lim_{\epsilon\to 0}\int{dk} \, e^{-ik^2t+k(y- ix) -\epsilon k^2},
      \label{fourier4} \\
   &= e^{-i\pi/ 4} \sqrt {m\over 2\pi t}\, e^{-im(y - ix)^2 / 2t}, \nonumber
\end{align}
and (\ref{fourier4}) extends analytically to the lower half plane.  

We have found the structure:
\begin{itemize}
\item
The function $-i\bra{y}e^{-iHt}\ket{x=0}$, calculated in the definite
   inner product representation, has a negative-energy
  spectral representation that 
  defines a function analytic
  in $t$ on the upper half plane.
\item
The function $\bbra{y}e^{-iHt}\kket{x=0}$, calculated in the indefinite
   inner product representation, 
has a positive-energy spectral representation that 
  defines a function analytic in $t$ on the upper half plane.
\item
On the positive real axis in $t$, these two functions coincide.  
Together, they therefore
define a single function analytic on the entire complex plane 
except for a cut on the negative real axis, where the value of 
$\bbra{y}e^{-iHt}\kket{x=0}$
is obtained by approaching the cut from below, and 
$-i\bra{y}e^{-iHt}\ket{x=0}$ is obtained by approaching the cut from above.    
\end{itemize}
Again, for positive real $t$, the results 
calculated using the two representations coincide.

\section{General potentials}
\label{general}

We now generalize the above analysis to a wider class of potentials. 
We shall see that the transition functions in the 
definite and indefinite quantizations typically
describe a single analytic function for different regions of 
complex time that overlap for an initial interval of real time.
In section \ref{ground}, we will apply this correspondence to 
the extraction of ground state correlation functions.    

The general problem will be as follows.  Consider a 
one-dimensional Hamiltonian 
$$
H \equiv - {p^2\over 2m} + \mu p + V(q).
$$
that is Hermitian with respect to the ordinary Hilbert space 
inner product on the real line, which requires that $\mu$ be real
and that $V(x)$ be a 
real-valued function of $x$.
Our discussion will include 
the case that is of interest in phantom and ghost theories, where
the spectrum of $H$, which is real in the ordinary Hilbert space 
representation, may be unbounded below.  As a result, the usual rotation to Euclidean time 
may not exist,\footnote{Although we shall see that the relevant unbounded
Euclidean path integral may in fact be calculable as a Gel'fand-Shilov
generalized function.} but typically there is no obstacle to calculating
the real-time transition function, which will be denoted by 
\begin{align}
  \bra{y}e^{-iHt}\ket{x}.   \label{propH}
\end{align}
In the indefinite inner product representation constructed in 
section \ref{freesect}, where the configuration space of $q$ is
the imaginary axis, 
the operators $p$ and $q$ are still Hermitian, as is the Hamiltonian.
However, the spectrum of the Hamiltonian may be complex.\footnote{We
remind the reader  Hermitian operators
in indefinite inner product spaces may have 
complex eigenvalues. These always occur in complex conjugate 
pairs.  The corresponding eigenstates are dual null states \cite{bognar}.}   
Of most interest will be those phantom or ghost theories for which,
in the indefinite representation, the 
real part of the 
spectrum becomes bounded below.\footnote{More
generally, our arguments remain valid even in cases where the real part 
of the spectrum is unbounded below, as long as the density of 
generalized eigenvalues decreases sufficiently rapidly in the negative
real direction.}   We will assume
that the imaginary part of the spectrum, if any, is sufficiently well-behaved 
that
$e^{-iHt}$ exists for all real $t$.  The transition function in the indefinite
representation is denoted by
\begin{align}
  \bbra{iy}e^{-iHt}\kket{-ix}.  \label{propK}
\end{align}
Due to the different configuration spaces, the quantities 
(\ref{propH}) and (\ref{propK}) are not directly comparable. 
For a comparison, we need states of the form 
$
  \bbra{y}
$
instead of $\bbra{iy}$, whereas general states in the indefinite
representation are functions $\phi(iy)\equiv \bbraket{iy}{\phi}$ that are
only defined on the imaginary axis.   Generalized functions such as 
 $\bbra{y}$ can, however,  be defined on 
suitable test function spaces that are invariant under time evolution
and consist of functions that may be 
analytically continued away from the imaginary axis 
into the complex plane.  The choice of test function space is
dependent on the Hamiltonian.  

Gel'fand and Shilov 
\cite{gelfand1, gelfand2} introduced families
$S_\alpha^\beta(x)$ of test function spaces on 
the real line that are suitable for 
our purposes.  These are defined, for $\alpha, \beta \le 0$,
 as consisting of all infinitely differentiable functions $\phi(x)$ on the 
real line satisfying the growth conditions
$$
  |x^k\phi^{(q)}(x)| \le C A^k B^q k^{\alpha k} q^{\beta q},
$$
where $A$, $B$ and $C$ may depend on $\phi$.   
Writing the independent variable $x$ in parentheses 
is not part of the 
original notation but will be necessary for later disambiguation
when we discuss similar spaces based on functions $\phi(ix)$
defined on the imaginary axis.  
There exists an interpretation of the limit of 
infinite $\alpha$ or $\beta$, and it is conventional to omit either 
index when it is  infinite. 
The familiar Schwartz space is the same as
$S = S_\infty^\infty(x)$.   
 For other values of $\alpha$ and $\beta$, the space is
more restricted than Schwartz space since
the  growth conditions are stricter, and in fact for $\beta \le 1$ these 
conditions are so strict that the functions $\phi(x)$ can be 
extended to analytic functions $\phi(x + iy)$ on some complex 
domain.   

The domain depends on $\beta$.
Specifically, for $\beta = 1$, the test functions $\phi(x)$ may be 
analytically extended to analytic functions on a strip 
$$
\{x + iy | x \in \mathbf{R}, |y| \le B\},
$$
where the width $B$ of the strip may depend on $\phi$.
For $\beta < 1$, the test functions
can in fact be continued  an entire 
analytic function on the whole complex plane.  In this case, the spaces 
$S_\alpha^\beta(x)$ are 
characterized completely by the growth conditions
\begin{align}
  |x^k \phi(x + iy)| \le C_k \, \exp \left({-a|x|^{1/\alpha} + b|y|^{1/( 1-\beta)}}\right), \label{growthab}
\end{align}
where  $C_k$, $\alpha$, and $b$ are positive real numbers that are
allowed to depend on $\phi$.  Smaller $\alpha$ or $\beta$ correspond to 
smaller spaces of test functions, and therefore larger dual spaces of
generalized functions.  The restrictions become so stringent that 
the test spaces are trivial for 
$$
  \alpha + \beta < 1.
$$
Under the Fourier transform,
it may be shown that \cite{gelfand1}
\begin{align}
  \mathcal{F}\left(S_\alpha^\beta(x)\right) = S^\alpha_\beta(k). \label{Sfourier}
\end{align} 
In the indefinite representation, the wave functions are not defined
on the real line but rather on the imaginary 
axis, and we will denote the spaces where the same
growth conditions are imposed in the imaginary direction 
by $S_\alpha^\beta(ix)$.
As above, for  $\beta<1$ these test functions 
can be continued away from the imaginary axis to obtain entire functions. 
For $\beta = 1$, the strip of analyticity will now be a vertical one
that includes the whole imaginary axis.  

The 
relation (\ref{Sfourier}) remains valid if we define the 
Fourier transform on the imaginary axis as 
$$
  \mathcal{F} [\phi] (ik) = \int dx\, e^{-ikx} \phi(ix).
$$
We will restrict our attention to potentials for which a test space
can be selected on which
the Trotter product formula
$$
  e^{-iHt} = \lim_{N\to \infty} \left(e^{-i\epsilon {p^2/ 2m} + i\epsilon\mu p}e^{-i\epsilon V}\right)^N,\quad \epsilon \equiv {t\over N},
$$
holds at least for a finite time interval.  This formula can
be taken a basis for the construction of a path integral representation.
If it holds,
 a generalized path integral 
may be defined by writing the (generalized) transition 
function $\bbra {y} e^{-iHt}$ as
\begin{align}
  \lim_{N\to\infty}\bbra {y}  \left(\mathcal {F}^{-1}\cdot e^{-i\epsilon {p^2/ 2m} + i\epsilon\mu p}\cdot \mathcal {F}\cdot e^{-i\epsilon V}\right)^N. \label {PI}
\end{align}
If one writes
 the Fourier transforms, each acting to the left on a generalized function,
 as formal integrals,
this can be seen to coincide with the
introductory textbook definition of the path integral.\footnote{It
should be remarked that even the simplest textbook real-time path integral
calculations give meaning to the oscillatory integrals 
involved in a way that is essentially equivalent to 
calculating Fourier transforms of distributions as we do here.
By using smaller invariant test spaces, our 
spaces of distributions are larger, allowing us to 
calculate a larger class of path integrals.}  For related definitions
of path integrals in terms of transforms, see for example \cite{dewitt} and 
the review in \cite{lapidus}.    

This expression will exist as a generalized function, but may not have 
a representation as an ordinary kernel.  In other words 
$$\bbra{y}e^{-iHt}\kket{-ix}$$ may not exist as an ordinary function
of $y$ and $ix$.  It is only when the kernel can be represented 
by a function continuous in $y$ and in $x$ at  $x = 0$ that we can compare
$$\bbra{y}e^{-iHt}\kket{-ix}$$ and $$\bra{y}e^{-iHt}\ket{x},$$ 
in the point $x = 0$.  Otherwise, as distributions, the objects
$$\bbra{y}e^{-iHt}$$ and $$\bra{y}e^{-iHt}$$ are incomparable, since they
act on different test spaces.

For the above construction, we saw that 
it is necessary to consider test spaces
consisting of analytic functions that are
invariant under all of $e^{-i \epsilon V(q)}$, $e^{-i\epsilon p^2/2m}$
and $e^{i\epsilon\mu p}$.

The appropriate test spaces will depend on the potential.  We will choose
suitable candidates from    
the family of spaces  
$S_\alpha^\beta(ix)$, whose elements are either entire 
functions for $\beta < 1$, or
analytic functions 
in a vertical strip including the imaginary axis for $\beta =1$.

We start with the spaces of entire functions 
$S_\alpha^\beta(ix)$ for $\beta < 1$.  These spaces will be suitable 
only for potentials such that $e^{-i \epsilon V(ix)}$ can be extended to 
an entire function 
of $z$, since the space must be closed under this multiplier.

The indices $\alpha$ and $\beta$ are then further restricted  by the growth 
properties of  $e^{-i \epsilon V(z)}$.  
We will consider the class of potentials satisfying the inequality
\begin{align}
  |e^{-i \epsilon V(z)}| \le C\, e^{b|z|^p}, \label{expgrowth}
\end{align}
on the complex plane, 
which is true in particular if $V$ is a polynomial of
order $p$.  Then a sufficient condition for $S_\alpha^\beta(ix)$ to 
be closed under this multiplier 
is that $p \le 1/\alpha$ and $p \le 1/(1-\beta)$, 
or\footnote{When the inequality is saturated, the space 
$S_\alpha^\beta$ is not necessarily
closed under the action of the multiplier.  In this case, 
the appropriate test function spaces are families of
 subspaces$S_{\alpha,A}^{\beta, B}$ of $S_\alpha^\beta$.   As noted in a previous footnote, this 
technical complication does not affect the general argument and 
will be glossed over in this article. For full details on the technique,
see
\cite{gelfand2}.}
\begin{align}
  \alpha \le {1\over p},\quad \beta \ge {p- 1\over p}.  \label{cond1}
\end{align}
Next, consider invariance of these spaces $S_\alpha^\beta(ix)$, still for $\beta<1$, 
under
$e^{\pm i\epsilon p^2/2m + i\mu p}$, which is equivalent to 
invariance of the Fourier transformed space $S^\alpha_\beta(ik)$
under
$e^{\pm i\epsilon (ik)^2/2m + i\epsilon\mu\,(ik)}$.  We will restrict attention to the 
family (\ref{cond1})
selected by the growth condition 
(\ref{expgrowth}).  First, if the order $p$ of the potential 
satisfies $p > 1$, then the first inequality constrains   
 $\alpha \le 1/p < 1$, so that 
the Fourier-transformed space also consists of entire functions. 
This space will be invariant if 
\begin{align}
  \beta \le {1\over 2},\quad  \alpha \ge {1\over 2}.  \label{cond2}
\end{align}
Solutions to both sets of conditions  (\ref{cond1}) and 
(\ref{cond2}) may be found
for potentials of 
order $1 < p \le 2$.  We may choose any such $\alpha$ and $\beta$ satisfying 
the additional condition $\alpha + \beta \ge 1$ necessary for
non-triviality of the test space.   
The smallest test space satisfying all these conditions for 
all $p$ in this
range is given by $S_{1/2}^{1/2}(ix)$, which happens to also
have  the nice property
of being invariant under the Fourier transform.
In fact, it is easily seen that $S_{1/2}^{1/2}(ix)$ satisfies 
the conditions for the extended range
$0\le p\le 2$, so that we may use as a test space for the whole range
$$
  S_{1/2}^{1/2}(ix), \quad 0 \le p \le 2.  
$$
In fact, the invariance conditions are unchanged 
for arbitrary complex 
coefficients of 
$p$, $p^2$, $x$ and $x^2$ in the Hamiltonian, enabling us to 
calculate path integrals as Gel'fand-Shilov distributions
on $
  S_{1/2}^{1/2}(ix)  
$ even when the integrand may be 
exponentially unbounded.  

We note that (\ref{cond1}) and (\ref{cond2}) cannot be satisfied
for polynomial potentials of order $p>2$.  As a result, none of the
Gel'fand-Shilov spaces of 
entire functions may be used for these potentials.  However, 
for our later arguments, it will be good enough to have test functions
that are analytic in a vertical strip containing the imaginary axis, 
a property satisfied by the spaces $S^1_\alpha(ix)$.  
This family of spaces, with suitable restrictions on $\alpha$, will
prove to be suitable for even potentials of order $p>2$.

According to \cite{gelfand1}, a sufficient condition for a 
function $f(ix)$, defined on the imaginary
axis, to be a multiplier with respect to  
a specific family of subspaces $S_\alpha^{\beta,B}(ix)$ of
$S_\alpha^{\beta}(ix)$, is
that it satisfy the estimate
\begin{align}
  |f^{(q)} (ix)| \le C_\epsilon\, B^q\, q^{\beta q}\, e^{\epsilon |x|^{1/\alpha}}
, \label {estimate}
\end{align}
for any $\epsilon > 0$.  
Consider, then, the term $V(q) = cq^p$ in a polynomial potential, 
where $c$ is real.  Its 
 contribution to the multiplier $e^{iV(ix)}$ is
$$
f(ix) = e^{i^{p+1} cx^p}.  
$$
We see that $p$ must be even to satisfy the sufficient condition above.  
By writing down the first 
few derivatives, it is easily seen  that 
$$
  |f^{q}(ix) \le q!\, p^q \left(1 + |x|^{pq}\right).
$$ 
This is not the tightest possible estimate, but it will suffice for 
our purposes.  Using the Stirling estimate for large q
$$
  \ln \left(q!\right) \sim q\ln q - q,
$$
we find 
\begin{align*}
  |f^{q}(ix)| &\le C\, q^q\,e^{-q}\, p^q \left(1 + |x|^{pq}\right) \\
  &\le C_\epsilon q^q\, p^q \,e^{\epsilon |x|^{1/\alpha}}.
\end{align*}
Comparing with the estimate (\ref{estimate}), we read off that any
$$
  \beta \ge 1, \quad \alpha >0,
$$
will do, for any even power $p$ in the potential. 
Of these, only $\beta=1$ gives a space of analytic functions.  
Applying the same argument to the 
multiplier $e^{\pm i\epsilon (ik)^2/2m}$ on the Fourier transformed
space $S^\alpha_\beta$ gives the (sufficient) condition
$$
  \beta > 0, \quad \alpha \ge 1.
$$
The smallest space of analytic functions satisfying both these conditions is
$$
  S_1^1 (ix),
$$
which, again, has the nice property of invariance under the Fourier transform.

In fact, while the above conditions are sufficient, it may easily 
be seen that the linear potential or source term 
$e^{i \lambda z}$ is also a multiplier on families of subspaces 
of $S_1^1(ix)$, \textit{despite} its exponential growth in the 
imaginary direction if $\lambda$ is real, so that we may allow 
also source terms in 
our potential.  This is not true for higher odd powers.  

In fact, it may be checked that both test spaces that we have chosen are
invariant under general source terms of the form
$$
  \mu p + \lambda q,
$$
where $\mu$ and $\lambda$ may be \textit{complex}.  

To summarize, we find that the following test spaces are suitable:
\begin{itemize}
\item
  $H = \nu p^2/2m + \mu p + V(q)$, where $\mu, \nu$ and $V(q)$ may be 
complex, as long as
   $|e^{-i \epsilon V(z)}| \le C\, e^{b|z|^p}$ for $0\le p \le 2$: A suitable
 test
space is $S_{1/2}^{1/2}(ix)$.  Elements of the test space are entire functions. 
\item
 $H = \pm p^2/2m + \mu p + \lambda q + V(q)$, where $\mu$ and $\lambda$
may be complex and 
   $V(q)$ is an arbitrary even real polynomial: A suitable test space is
$S_1^1(ix)$.  Elements of the test space are analytic on 
a vertical strip including the
imaginary axis.  
\end{itemize}
A useful feature of the current formalism is that it allows
the study of source terms in phantom or ghost theories.  These caused some
difficulty in prior treatments \cite{BG, sakoda}.  
In particular, while the exponential of a
 source term  
$e^{i \lambda z}$ blows up on the imaginary axis where the 
states of the ghost theory are defined, it is still a legitimate
multiplier on $S_{1/2}^{1/2}(ix)$, where its exponential growth is 
dominated by the faster rate of decrease of the test functions,
and on families of subspaces of
$S_1^1(ix)$, where it just modifies the coefficient, and not the 
exponential power, of decrease of test functions.   

\textit{In both cases, 
the path integral may be defined, at least for a finite initial time
interval,  via the formula
(\ref{PI}), which defines a legitimate distribution
 despite the fact that the integrand in the 
naive path integral may be exponentially unbounded.}  

A similar analysis can be made for the definite inner product 
version whose configuration space is the real line by replacing
$ix$ by $x$ in the arguments.

Potentials for which both $S_{1/2}^{1/2}(ix)$ and $S_1^1(ix)$ 
are suitable include the free particle
and the harmonic oscillator, as well as 
the linear potential, of interest in some cosmological models,
and the inverted harmonic potential that appears in two-dimensional string
theory, despite the fact that the potentials in the latter two cases
 are unbounded below. 
The path integral may be defined for 
these potentials via the formula (\ref{PI}).

While $S_1^1(ix)$ allows the treatment of higher
polynomial potentials that
cannot be treated using $S_{1/2}^{1/2}(ix)$, 
we note that 
 $S_{1/2}^{1/2}(ix)$ allows us to treat some
potentials that cannot be treated using $S_1^1(ix)$.  In particular, while
$S_1^1(ix)$ is good for real polynomials, the growth condition 
 $|e^{-i \epsilon V(z)}| \le C\, e^{b|z|^p}$ for  
$S_{1/2}^{1/2}(ix)$  does not require the momentum terms or
 $V(q)$ to be real.  

\textit{In this case, the expression (\ref{PI})
defines a good path integral even though the integrand
may exponentially unbounded with growth of order $e^{ak^2 + bx^2}$.}

Given a test space consisting of analytic functions, a generalized function
$\bbra{z}$ can be defined as 
$$
  \bbraket{z}{\phi} \equiv \phi(z),
$$
for any $z$ within the common domain of analyticity of the test functions 
 $\phi$,
even though the full space of states is defined only on the imaginary axis.  
 
We will consider the phantom Hamiltonian introduced at the 
beginning of the section
\begin{align}
H \equiv - {p^2\over 2m} + \mu p + V(q). \label{generalphantomH}
\end{align}
and argue that under certain assumptions the quantities
$$
  \bra {y} e^{-iHt} \ket{x=0}
$$
and 
$$
 i \bbra {y} e^{-iHt} \kket{x=0},
$$
coincide for some range of $y$ and for some initial time 
interval $0 < t < T$.
This will hold for any real $y$ 
when we use the test space $S_{1/2}^{1/2}(ix)$, whose elements are
entire functions, or for $y$ in a
neighbourhood of $0$  in the case of $S_1^1(ix)$, whose elements
are analytic only on a strip including the imaginary axis.

Our first assumption is that the distribution 
$\bbra {z} e^{-iHt}$ can be represented by a kernel,  denoted by 
$$
\bbra {z} e^{-iHt} \kket{-ix},
$$
that is a 
differentiable function in $\Re(z)$, $\Im(z)$ and $x$ and a differentiable function of
$t$ for 
some initial time interval $0 < t < T$.\footnote{The example of the
harmonic oscillator, for which the transition function
 refocuses to a delta distribution after a half
period, shows the need for the time interval restriction.}  Given this
assumption, we may write
$$
  \phi(z, t) \equiv
  \bbra {z} e^{-iHt} \kket{\phi} = \int dx\,\bbra {z} e^{-iHt} \kket{-ix}\, \phi(ix, 0)
$$
for test functions $\phi \in S_{1/2}^{1/2}(ix)$ or
$\phi \in S_{1}^{1}(ix)$.  Here
$\phi(z, t)$ is 
analytic, since the space of test functions 
was chosen to be closed under time evolution.
Differentiating on both sides with respect 
to $\partial_{\bar z}$, and using the fact that the spaces 
$S_\alpha^\beta(ix)$ are ``sufficiently rich''
\cite{gelfand1}, we conclude that
$$
  \partial_{\bar z} \bbra {z} e^{-iHt} \kket{-ix} = 0.
$$
In other words, $\bbra {z} e^{-iHt} \kket{ix}$ is analytic in $z$.  

Due to this analyticity of $\bbra {z} e^{-iHt} \kket{ix}$, the
differential equation
$$
  \left(i \, \partial_t - H \left(-\partial_y, iy\right)\right)
    \bbra {iy} e^{-iHt} \kket{x=0} = 0,
$$
which is satisfied by assumption 
for $0<t<T$ by the transition function in the indefinite representation
($q \to ix$, $p \to -\partial_x$) on the imaginary axis, is in fact 
true on the whole domain of analyticity in $z$.  We find
$$
  \left(i \partial_t - H \left(-i\partial_z, z\right)\right)
    \bbra {z} e^{-iHt} \kket{x=0} = 0,
$$
on the entire real
axis in the case of $S_{1/2}^{1/2}(iy)$, or some real neighbourhood of
zero in the case of
$S_{1}^{1}(iy)$.  But for real values of $z$,
 this just coincides with the ordinary
Schr\"odinger equation giving the time evolution of 
$\bra {y} e^{-iHt} \ket{x=0}$ in the ordinary Hilbert space representation.  

Therefore, on the interval $0<t<T$, the objects 
$$i\bbra {y} e^{-iHt} \kket{x=0}$$
and 
$$\bra {y} e^{-iHt} \ket{x=0}$$
are ordinary functions 
satisfying the the same differential equation in $y$.  
They would therefore coincide if they satisfied the same initial conditions.
However, the initial conditions at $t=0$ in the two representations are 
not ordinary functions, but rather
delta distributions on two different test spaces.  These 
distributions are 
not comparable.

We therefore recast the problem in a form that gives
an ordinary function as initial condition.  This is done by 
considering the quotient
$$ G(z, t) \equiv {\bbra {z} e^{-iHt} \kket{x=0} \over\bbra {z} e^{-iH_0 t} \kket{x=0}},$$
where $H_0$ is the free phantom Hamiltonian considered in 
section \ref{phantomfree}, and the denominator is 
the analytic continuation to all $z\in\mathbf{C}$
\begin{align}
 \bbra {z} e^{-iH_0 t} \kket{x=0} = e^{-i\pi/4}
    \sqrt {m\over 2\pi t} \, e^{-imz^2/2t}.  \label{G0}
\end{align}
of the transition function 
$$
  \bbra {iy} e^{-iH_0 t} \kket{x= 0} = e^{-i\pi/4}
    \sqrt {m\over 2\pi t} \, e^{imy^2/2t} \to \delta (y) \quad {\textrm{as}}
  \quad t\to 0,
$$
of the phantom particle on its imaginary axis configuration space.
Given the form (\ref{generalphantomH}) of the Hamiltonian, 
the differential equation satisfied by $G(z, t)$
is 
\begin{align}
 \left(i t \partial_t - tH \left(-i\partial_z, z\right)
    +iz\partial_z - \mu m z\right)
  G(z, t)  = 0,   \label {indefeq}
\end{align}
for $0 < t < T$, with initial condition the ordinary function
$$
  G(iy, t) \to 1, \quad t\to 0,
$$
 on the imaginary axis, assuming
that the effect of the interaction becomes negligible 
in the limit as $t\to 0$, so that the behaviour of the numerator 
approaches that of the denominator in this limit.
For the class of smooth potentials under discussion, we 
expect this to hold, although we do not have general proof.\footnote{It
is physically reasonable that the potential requires time to make
its presence felt, so that the behaviour approaches that of
a free particle in the zero time limit.}

To compare this solution with the one obtained in the 
definite representation, we need to
know what happens to $G(z, t)$ for real values of $z$ as $t\to 0$.  To this end, we
note that the solution of this differential equation gives a family 
of functions, indexed by $t$, that are
 analytic on an open domain in the complex plane in $z$ including
either a vertical strip containing the imaginary axis or
the whole complex plane, depending on the choice of test space.  
This family approaches 
to the constant function $1$ on the imaginary axis.  
If this family is uniformly bounded on some
open neighbourhood of the origin for small
enough $t$, we  
may use Vitali's theorem \cite{vitali1, vitali2} to
conclude that the family converges to the constant function 
 $1$ not only on the imaginary axis but 
on this whole neighbourhood and, by implication, also on the 
segment
of the real axis included in the neighbourhood.  

For the class of potentials under discussion, we expect the condition of 
uniform boundedness to be true, although we have not found a proof.  
We will in the 
following assume this assertion without proof.  

We would like to compare $G(y, t)$ to the corresponding quantity 
$$
\tilde G(y, t) \equiv {{\bra {x} e^{-iHt} \ket{x=0} \over \bra {y} e^{-iH_0 t} \ket{x=0}}},
$$
in the
definite representation, 
which satisfies the differential equation 
\begin{align}
 \left(it \partial_t -t H \left(-i\partial_y, y\right)
    +iy\partial_y - \mu m y\right)
  \tilde G(y, t) = 0,   \label {defeq}
\end{align}
for $0 < t < T$, with the initial condition
$$
\tilde G(y, t) \to 1, \quad t\to 0.  
$$
Since the equations (\ref{indefeq}) and (\ref{defeq}) coincide for real
$z$  and have the same initial  
condition on the real line, their solutions coincide on the 
indicated time range.  For this range, we conclude 
\begin{align}
   \tilde G(y, t) = G(y, t),  \label {eql}
\end{align}
From their explicit expressions, 
the free transition functions in the two representations can be compared to obtain
\begin{align}
  \bbra y e^{-iH_0 t} \kket {x=0} & =  e^{-i\pi/4}
    \sqrt {m\over 2\pi t} \, e^{-imy^2/2t} \nonumber\\
 &= -i \bra y e^{-iH_0 t} \ket {x=0}, \quad t>0, \label{GOfree}
\end{align}  
and from (\ref{eql}) we find the result
\begin{align}
   \bra {y} e^{-iHt} \ket{x=0} = i\bbra {y} e^{-iHt} \kket{x=0},
\quad 0< t < T.  \label{main}
\end{align}
This is the main result allowing us to relate quantities calculated 
in the definite and indefinite representations.  
It states that, for an initial time interval,
certain transition functions calculated
in the definite and indefinite representations coincide up to a 
phase factor.  The upper limit 
of coincidence $T$ is the leftmost point on the positive time axis where the
transition function becomes singular (a proper distribution not 
representable by an ordinary function of $y$).

We now discuss the structure of the transition functions $\bra {y} e^{-iHt} \ket{x=0}$
and $\bbra {y} e^{-iHt} \kket{x=0}$ as functions of $t$
extended to the complex plane.  The above results do not depend 
on any bounds on the real part of the spectrum of $H$ in either representation.
As noted in the introduction to this section, however, 
the most interesting case occurs when the real part of the
spectrum in the indefinite representation is bounded below (or if not, 
the spectral density decreases sufficiently fast) so that 
$$
  \bbra y e^{-iH t} \kket {x=0}
$$
can be analytically extended to the lower half plane in $t$.  We will
discuss this case.\footnote{The discussion for other permutations of 
conditions on the definite or indefinite spectra would proceed
similarly.}   
 
Since we are not assuming any conditions on the spectrum of $H$ in the 
definite representation, the quantity
$$
  \bra y e^{-iH t} \ket {x=0},
$$
is not necessarily the boundary value of an analytic function in 
$t$ on the range of $t$ for which it exists.\footnote{
For an example where this function is not the boundary value
of an analytic function, see section \ref{combining}.}
However, since 
$$
i \bbra {y} e^{-iHt} \kket{x=0} = \bra {y} e^{-iHt} \ket{x=0},
$$ 
on an initial interval $0<t<T$, we may recover by
analytic continuation the full
$$
  \bbra y e^{-iH t} \kket {x=0}
$$
for all $t$ in the lower half of the complex plane from 
the values of $\bra {y} e^{-iHt} \ket{x=0}$ on this initial interval.  

In specific cases, the spectrum of $H$ may, in addition, 
be bounded above in the definite representation.
This is true for the free phantom particle and the phantom harmonic 
oscillator, but not of the two-particle  system described in section 
\ref{combining}.  In this case, the function 
$$
 \bra y e^{-iH t} \ket {x=0}
$$
can be analytically extended to the upper half plane in $t$.  Together 
with
 $$
i \bbra y e^{-iH t} \kket {x=0},
$$
it therefore defines a single analytic function  on the 
complex plane, with possible singularities on the real axis. 
The value for real $t$ of 
$ \bra y e^{-iH t} \ket {x=0}$ is obtained by approaching the real axis
from above, while the value of $
i \bbra y e^{-iH t} \kket {x=0},
$ is obtained by approaching the real axis from below.  These singularities
may, among other possibilities,
 take the form of branch points and cuts, as some of the examples in this
article illustrate.  

There will always be at least one singularity  on the real axis at 
the origin $t = 0$, where the transition function approaches the delta distribution.  
Repeating the analysis of the differential equation for 
negative times, and using the fact that the 
free phantom particle transition function for $t<0$ requires the opposite sign on 
the right of (\ref{GOfree}), we see that
$i\bbra {y} e^{-iHt} \kket{x=0}$ will instead coincide
with $- \bra {y} e^{-iHt} \ket{x=0}$ for small negative $t$.  
As a result, in the special case where there are
cuts on the real axis, the origin will be a branch point with the 
corresponding branch cut extending to the left.  

In general, there will be a singularity in $t$
wherever the solution becomes a proper distribution not representable
by an ordinary function.
For example, the harmonic oscillator, due to the common periodicity
of all the modes,
will have singularities in $t$ at all multiples of the half period.
In this case, the upper limit $T$ of the range of validity of
\begin{align}
   \bra {y} e^{-iHt} \ket{x=0} = i\bbra {y} e^{-iHt} \kket{x=0},
\quad 0< t < T.  \label{main1}
\end{align}
is one half of the period.  
The fact that there is such an upper limit appears to be a special
consequence of the even spacing of the harmonic oscillator energy levels,
which causes the wave function to be refocused to a delta function
after a half period.  
In general, perturbations that are not linear or harmonic are expected 
to destroy this even spacing, erasing the singularity that forces us
to take $T$ finite.  Therefore, for generic potentials, except at exceptional 
values of the constants in the potential, the period 
may be expected to become infinite, so that $T$ will be 
infinite and the above result should be valid for all positive $t$.

\section{Ground state expectation values}
\label{ground}

In this section we will argue that ground state 
expectation values appropriate to the 
indefinite representation can be extracted from the corresponding 
calculation in the definite representation.  

Since a theory is completely determined
by its ground state expectation values, one may conclude that
the definite representation encodes all the physical information
relevant to the indefinite representation, so that we can use
whichever representation is most convenient to perform 
calculations.\footnote{
The definite representation may 
be preferred for functional integral calculations in field theories, 
since the 
functional integral obtained from the definite representation
is in many cases manifestly covariant (modulo boundary conditions), whereas
the functional integral obtained from the indefinite representation 
is not manifestly covariant due to 
an unusual integration region for the phantom components of the 
fields \cite{BG, arisue}.  
Examples where this happens
include the time-like component of a gauge field \cite{sakoda}    
and the Dirac boson.   Dirac bosons appear
in Pauli-Villars regularizations of field theories containing fermions
and in Lee-Wick
field theories, and are discussed
in section \ref{dirac}.  
The results of this section may be seen as a justification for 
the covariant treatment of these fields in the functional integral. }

As in section \ref{general}, we will discuss
the case that is of most interest in phantom and ghost theories, where
the spectrum of $H$, which is real in the definite representation, 
may be unbounded
above and below.  
In the indefinite representation, the spectrum may be 
complex \cite{bognar}.  In the cases of interest to us, the real part 
of the indefinite representation 
spectrum will be bounded below, which we will assume.  We 
also assume that the indefinite representation has a unique state, called the 
ground state, for which the
real part of the energy spectrum attains this lowest bound.

We will now argue that, under these assumptions,
either representation may be used to calculate the
ground state expectation values relevant to the 
indefinite representation.

General expectation values may be generated 
by adding a time-dependent source term to the action.  
The argument is then based on a generalization of the result
of section \ref{general},
whose arguments and conclusions
remain valid, with slight modifications,
 if the Hamiltonian contains time-dependent linear source terms of the form
$$
  \mu (t)\, p + \lambda(t) \, q,
$$
where $\mu$ and $\lambda$ are real-valued functions of $t$.
   We already know that the relevant 
 test spaces admit time-independent linear terms of this type.
However, the argument determining the 
choice of test space was based on the space being closed 
under time evolution in the limit of vanishingly
small time increments, and therefore remains equally valid
when the source terms are allowed to depend on time.  

The further modifications needed are as follows:  Since
$H$ in (\ref{indefeq}) may now depend on time, we  replace
$$
  e^{-iHt} \to U(t_f, t_i)
$$ 
in (\ref{indefeq}) and subsequently.  The correspondingly 
modified conclusion (\ref{main1})
of section \ref{general} may now be written as
\begin{align}
  i\bbra{y} U(t_f, t_i) \kket{x=0} =  \bra{y} U(t_f, t_i) \ket{x=0},
   \quad -T < t_i < t_f < T. \label{withsource}
\end{align}
By taking functional derivatives with respect to $\mu(t)$ and
$\lambda(t)$, we may 
generate expectation values of insertions of
 powers of $p$ and $q$ in the two 
representations.  These will
then coincide up to a factor $i$ if all insertions are in the region $-T<t<T$ 
where (\ref{withsource}) is valid.  

For simplicity, we specialize our argument to the two-point function
of $q$.
Similar arguments can be made for other correlation functions.  
The above result implies that
\begin{align}
&\bbra{y, T} T q(t_f)\, q(t_i) \kket{x=0, -T} \nonumber \\ 
&\quad =
  \bbra{y} e^{-iH(T - t_f)}\, q\, e^{-iH(t_f-t_i)}\, q\, e^{-iH(t_i + T)}
   \kket{x=0} \nonumber\\
  &\quad = -i  \bra{y} e^{-iH(T - t_f)}\, q\, e^{-iH(t_f-t_i)}\, q\, e^{-iH(t_i + T)}
   \ket{x=0} \label{coincidence}
\end{align} 
provided we choose $T$ small enough and provided that 
$$
  -T < t_i <t_f < T.
$$
First, we would like to get rid of the 
restriction on the range of $t_i$ and $t_f$ to obtain a useful relation 
also outside the interval $(-T, T)$.  Second, we would like to extract the
ground-state expectation value from either representation.  

Although we do not know of a general argument to get rid of the 
restriction on the range of $t_i$ and $t_f$, we may do so in specific cases
by one of at least two methods.

First, as we remarked in section \ref{general}, for generic non-harmonic
potentials one expects $T$ to be infinite.  In this case, the 
restriction on $t_i$ and $t_f$ disappears and we are done.  

Second, in the cases where $T$ cannot be chosen infinite,
the expressions
$$
   \bbra{y} e^{-iH(T - t_f)}\, q\, e^{-iH(t_f-t_i)}\, q\, e^{-iH(t_i + T)}
   \kket{x=0},
$$  
and 
$$
   -i  \bra{y} e^{-iH(T - t_f)}\, q\, e^{-iH(t_f-t_i)}\, q\, e^{-iH(t_i + T)}
   \ket{x=0},
$$
generally do define functions of $t_f$ and $t_i$ also 
outside the range $(-T, T)$.  Although now these expressions have
 singularities on the real axis at finite values of $T$,
in various cases of interest they 
do describe analytic functions of $t_f$ and $t_i$ on the entire real 
line.\footnote{See 
for example the phantom harmonic oscillator of section \ref{pharmonic}.} 
Indeed, there are often separate considerations from which analyticity
of the two-point functions may be inferred.  
Since we have shown that these expressions 
coincide for sufficiently  small $t_f$ and $t_i$, in the case where they
define analytic functions they will coincide on  the 
entire region of
analyticity in $t_i$ and $t_f$.  

When a ground state exists in the indefinite representation
and has overlap with the state $\kket{x=0}$, the ground state 
expectation value may be 
extracted in the standard way by writing
\begin{align*}
   &\bbra{x=0} e^{-iH(T - t_f)}\, q\, e^{-iH(t_f-t_i)}\, q\, e^{-iH(t_i + T)}
   \kket{x=0} \\
&\quad = \sum_{m, n, k}\bbraket{x=0}{m} e^{-iE_m(T - t_f)}\,\bbraket{m}{m} 
\bbra{m}q\kket{n} e^{-iE_n(t_f-t_i)}\bbraket{n}{n}\times \\
   &\quad\quad\qquad \times \bbra {n}q\kket{k} e^{-iE_k(t_i + T)}
   \bbraket{k}{k}\bbraket{k}{x=0},
\end{align*}  
and
\begin{align*}
  &\bbra{x=0} e^{-iH(2T)}\kket{x=0} = 
   \sum_m \bbraket{x=0}{m} e^{-iE_m(2T)} \bbraket{m}{m} \bbraket{m}{x=0},
\end{align*}
and taking the limit $T\to -i\infty$ of the analytic continuation in $T$ of
their quotient
to the lower half plane.\footnote{We have 
assumed here for notational convenience
that the spectra are discrete and do not contain zero norm 
states, so that a basis can be chosen satisfying $\bbraket{m}{n} = \pm \delta_{mn}$.  The generalizations when this is not the case are straightforward.}
  This analytic
continuation exists given the bound  on the indefinite 
representation spectrum that we assumed in the introduction to this 
section.\footnote{
It is important to note that in general we have to take $x=0$ in
the final state
 $\bbra{x=0}$,
since the eigenstates $\kket{m}$ do not necessarily belong to the  
subspace of analytically continuable test functions.  Since the
indefinite representation configuration
space is the imaginary axis, $\bbraket{x}{m}$ is only guaranteed to 
be meaningful for $x = 0$. 
This restriction on the final boundary condition
may be relaxed in cases where all the eigenstates $\kket{m}$ belong to the 
test space.  }

Denoting the ground state by $\kket{0}$, we obtain
\begin{align}
  \bbra {0} q(t_f) \, q(t_i) \kket{0} = \lim_{T\to -i\infty}
{\bbra{x=0} e^{-iH(T - t_f)}\, q\, e^{-iH(t_f-t_i)}\, q\, e^{-iH(t_i + T)}
   \kket{x=0} 
\over \bbra{x=0} e^{-iH(2T)}\kket{x=0}
}.  \label{gsindef}
\end{align}
By (\ref{main1}) and (\ref{coincidence}), 
we know that the definite and indefinite 
representations of the numerator and denominator on the right hand side
 coincide up to a common factor $i$ for sufficiently small real $T$, and will
therefore have the same analytic continuation to the lower half plane in $T$.  
We may therefore just as well write
\begin{align}
  \bbra {0} q(t_f) \, q(t_i) \kket{0} = \lim_{T\to -i\infty}
{\bra{x=0} e^{-iH(T - t_f)}\, q\, e^{-iH(t_f-t_i)}\, q\, e^{-iH(t_i + T)}
   \ket{x=0} 
\over \bra{x=0} e^{-iH(2T)}\ket{x=0}
}.\label{gsdef}
\end{align}
\textit{This formula expresses the ground state 
expectation value relevant to the indefinite representation
in terms of quantities calculated in the definite representation, 
which may not have a ground state.}

To make the connection with the path integral, we note that it is only when
 $-T < t_i, t_f < T$ that the expressions
have a simple path integral representation. 
With this provision, we may express the result as 
\begin{align}
 & \bbra {0} T q(t_f) \, q(t_i) \kket{0} \nonumber \\ 
&\qquad= \lim_{T\to -i\infty}
{\bra{x=0, T} T q(t_f)\,q(t_i)\ket{x=0, -T} 
\over \braket{x=0, T}{x=0, -T}}, \quad -T < t_i < t_f < T.
  \label {expectresult}
\end{align}
We call the reader's attention to the fact that this formula
enables us to extract the 
indefinite ground state expectation value from the 
path integral obtained from the definite representation,
where there is no ground state.

Once again, we repeat that for sufficiently generic potentials we expect $T$ to be
infinite, so that there is no need to restrict the ranges of 
$t_i$ and $t_f$.  When $T$ is finite, the path integral calculation may be 
done assuming $-T < t_i, t_f < T$, and the general two-point-function 
for $t_i$ and $t_f$ outside this range must in general be obtained from 
separate analyticity considerations in $t_i$ and $t_f$.  

Even if the upper limit for 
$T$ is finite, the restriction on $T$ may still be overcome even 
without such analyticity considerations
if the definite 
spectrum is bounded above, as in the case of the phantom harmonic 
oscillator  
discussed below.  When there is such a bound, 
the expectation values on the right hand sides of
 (\ref{gsindef}) and (\ref{gsdef})
describe the boundary values of analytic functions defined respectively on 
the
lower and upper half planes in $T$.  Since these coincide for small enough
$T$, they in fact describe a single analytic function with  
singularities or cuts on the real axis, but which can be analytically 
continued from the upper to the lower half plane past the above interval
of coincidence.    
This analytic function, and its limit for $T\to -i\infty$, can 
therefore be recovered from the value of the definite representation for 
any range of $T$.  

After this rather technical discussion, it may be helpful to 
remind the reader why the result  (\ref{expectresult}) is useful to us.  
As we remarked, in concrete applications 
the definite representation is often the one 
where covariance is manifest, a property that becomes obscured in the 
indefinite representation.   If we do this, the
absence of a ground state in the definite representation may make
a Euclidean time quantization problematic,
 whereas the 
real-time quantization considered in this paper
remains well-defined.  In a 
non-perturbative path integral approach, we do not know the ground state 
a priori, but the fixed boundary conditions in 
quantities such as
$$\bra{x=0, T} T q(t_f)\,q(t_i)\ket{x=0, -T}$$
occurring in  (\ref{expectresult}) are naturally and easily imposed.  

We note in conclusion that, since in both representations we have to 
perform an analytic continuation to extract the ground
state expectation values, the indefinite representation provides
no relative advantage compared to the definite representation for the 
calculation of these quantities.

\section{Inverting the construction}

So far we have argued that, for certain 
classes of  Hamiltonians, ground state 
expectation values in an indefinite inner product representation
can be obtained from transition amplitudes calculated in the
positive-definite Hilbert space quantization of the Hamiltonian.  

It is conceivable that, for some  Hamiltonians that do not 
belong to the classes we discussed, the transition amplitudes in the 
positive-definite quantization on a given initial time interval
may be analytically extensible to the lower half plane.  In this 
case, one would still expect this 
analytic continuation to display 
properties of the indefinite inner product representation.

Studying this would require developing the argument of the previous sections
in reverse, and is beyond the scope of this article.

\section{The phantom harmonic oscillator}
\label{pharmonic}

In this section we study
a  harmonic oscillator with opposite sign action.  
It is now the indefinite representation
that has positive energies and a ground state.
We show that ground state expectation values for the indefinite
representation may be 
extracted from the definite-representation path integral 
in the usual way, leading to an unambiguous $i0$ prescription
for the two-point function.  

Let us first review the canonical quantization of the 
phantom Lagrangian
\begin{align}
  \mathcal{L} = -\half \,m \dot x ^2 + {m\omega^2\over 2}\, x^2.
   \label{phantomL}
\end{align}
We read off that 
$$
  p = -m\dot x,
$$
and canonical quantization instructs us to take $[x, p] = i$.  Defining
as usual
\begin{align}
  a &= \sqrt{m\omega\over 2} \left( x + {i\over m\omega} \, p \right), \\
  a^\dagger &= \sqrt{m\omega\over 2} \left( x - {i\over m\omega} \, p \right), 
\end{align}
where by convention 
$$\omega>0,$$ we obtain
$$
  [a, a^\dagger] = 1  
$$
and the Hamiltonian is given by
$$
 H = -{p^2\over 2m} - {m\omega^2\over 2}\, x^2 =  - {\omega\over 2}  \left(a a^\dagger + a^\dagger a \right).
$$
Note that we would have obtained the same Hamiltonian if
we had started from 
the first-order form 
\begin{align}
  \mathcal{L} = {i\over 2}\left(a^\dagger \dot a - \dot a^\dagger a\right)
                + \omega\, a^\dagger a,
\label{phantomfo}
\end{align}
of the original Lagrangian.

We now have two choices of representation.  The first is the 
familiar positive inner product representation generated from 
a state $\ket 0$ satisfying
$$
  a\ket 0 = 0.  
$$
The resulting states 
$$\ket n \equiv {1\over \sqrt {n!}}\, (a^\dagger)^n\ket 0$$
 have positive inner products and
negative energies
$$
  E_n = -\omega\left(n + \half\right),\quad n = 0, 1, 2, \dots,
$$
and there is no ground state.  In position space, the 
normalized states are 
\begin{align}
  \braket {x}{n} = {1\over \sqrt {2^n\, n!}} 
\left({m\omega\over \pi}\right)^{1/4}\, H_n (\sqrt {m \omega}\,x )\,e^{-m \omega\, x^2/2}.  \label{posstates}
\end{align}
The other representation lives in the indefinite inner product space 
introduced in section \ref{freesect}, and is
generated from a state $\kket {0}$ satisfying 
$$
  a^\dagger \kket {0} = 0, \quad \bbraket{ 0}{ 0} = 1.
$$
For notational convenience, we define
\begin{align*}
  b &\equiv a^\dagger, 
\end{align*}
so that
\begin{align*}
   [b, b^\dagger] = -1, \quad
    b \kket{ 0}  = 0.
\end{align*}
An easy consequence of the negative sign in the commutation relation is
that the inner products of the 
states 
$$\kket {n} \equiv {1\over \sqrt {n!}}\,\left(b^\dagger\right)^n\kket {0}$$ 
alternate in sign, and that these 
states have positive
energy 
$$
  E_{n} = + \omega \left(n + \half\right),\quad n = 0, 1, 2, \dots,
$$
so that the state $\kket{0}$ is in this case a ground state.  From the 
differential equation implied by $b \kket {0} = 0$, we may
construct these states in position space.  We find that they are
\begin{align}
  \bbraket {ix}{ n} = {1\over \sqrt {2^n\, n!}} 
\left({m\omega\over \pi}\right)^{1/4}\, H_n (\sqrt {m \omega}\,(-x) )\,e^{-m \omega\, x^2/2}, \label{negstates}
\end{align}
on the imaginary axis where the wave functions take their values
according to the discussion of section \ref{freesect}. 
These are normalized states with respect to the 
indefinite inner product $\bbraket{\cdots}{\cdots}$ defined in
section \ref{freesect}.  In fact, we have 
$$
  \bbraket{m}{n} = (-)^m\, \delta_{mn}.
$$
This indeed coincides with the inner product implied by the commutation relations 
of $b$ and $b^\dagger$.  The completeness relation is 
\begin{align*}
   1 &= \sum_{n=0}^\infty \kket{n}\bbraket{n}{n} \bbra{n} \\
    &= \sum_{n=0}^\infty (-)^n \,\kket{n} \bbra{n}.
\end{align*}
We can straightforwardly evaluate the 
ground state expectation value
\begin{align}
   \eexpect{T b(t)\, b^\dagger (0)}
   &= - \theta (t)\,e^{-i\omega t}.
\label{twopoint0}
\end{align}
Expressing 
$$x = {1\over \sqrt {2m\omega}}\left(a + a^\dagger\right)
   = {1\over \sqrt {2m\omega}}\left(b^\dagger + b\right),$$ 
this implies
\begin{align}
   \eexpect{T x(t)\, x (0)} = - {1\over 2m\omega}\,e^{-i\omega |t|}.
\label{twopoint}
\end{align}
Note that the sign is opposite to that of the ordinary 
harmonic oscillator.

We now consider the path integral quantization of the theory.  
We start from the Hamiltonian formulation of the path integral
corresponding to the positive-definite inner product representation, 
which will enable us to carefully compute overall normalization 
factors important for the subsequent discussion.  We 
need to calculate
\begin{align}
 & \int [dp]\, [dx]\, \exp {i\int dt\, \left(p\dot x - H\right)} 
\label{fullPI} \\
   &\quad = \int [dp]\, [dx]\, \exp{i\int dt\, \left(p\dot x + {p^2\over 2m} 
      + {m\omega^2\over 2}\, x^2\right)} \nonumber\\
  &\quad \equiv \lim_{\Delta t \to 0} \int \prod_i {dp_i\over 2\pi}\,
          \prod_i{dx_i}\, \exp {i \sum_i \Delta t \left(p_i\,(x_{i+1} - x_i)
              + {p_i^2\over 2m} + {m\omega^2\over 2}\, x_i^2\right)} \nonumber\\
  &\quad = \lim_{\Delta t \to 0} 
    e^{i\pi/ 4}\,\sqrt {m\over 2\pi\, \Delta t}\int 
          \prod_i\left(e^{i\pi/ 4}\,\sqrt {m\over 2\pi\, \Delta t}\,{dx_i}\right)\nonumber\\
  &\qquad\qquad\qquad\qquad\qquad\qquad \times \exp {i \sum_i \Delta t \left(- \half\, m \left({x_{i+1} - x_i
   \over \Delta t}\right)^2 + {m\omega^2\over 2}\, x_i^2\right)},\nonumber
\end{align}
where we have integrated over the $p_i$ in the last step.  The 
prefactor comes from
the integration over the final $p_i$, while the final $x_i$ is fixed.

Note that the last line contains the expected discretization of the
Lagrangian $\mathcal{L}$ in equation (\ref{phantomL}).  However, it is 
important to note that the measure factors
$$
  e^{i\pi/ 4}\,\sqrt {m\over 2\pi\, \Delta t}
$$
are different from the ones that would have been obtained from the 
usual harmonic oscillator $-\mathcal{L}$, which has a 
factor $e^{-i\pi/ 4}$ instead of $ e^{i\pi/4}$.  

We will be interested in calculating ground state expectation 
values appropriate for the \textit{indefinite} inner product
representation, whereas
the above  path integral is the one we would obtain 
from the \textit{positive-definite}
representation.  However, in section \ref{ground} we argued that the 
former can be extracted from the latter via the usual
procedure of calculating the path integral for real 
time intervals $T$ and then taking the $T\to -i\infty$
limit of its analytic continuation, 
as long as we take the initial and final boundary conditions fixed at
$$
x_0 = x_N = 0,
$$
and provided these states have some overlap with the ground state.  
This is exactly the same procedure one would use to extract 
ground state expectation values, if one existed, for the
positive-definite inner product representation.\footnote{In other words,
the $T\to-i\infty$ limit automatically chooses whichever representation has a 
ground state.}

The change of variables
\begin{align}
  x_j = \sum_p q_p\,\sin {p\pi j \Delta t\over T} \label{change}
\end{align}
can be checked to have Jacobian
$$
  \sqrt {N \over 2}^{\phantom{i}N - 1},
$$
and we obtain
\begin{align*}
   &e^{i\pi/ 4} \lim_{N\to\infty} \sqrt{Nm\over 2\pi T}\,
     \prod_{p=1}^{N-1} \int dq_p\, e^{i\pi/ 4}\,\sqrt{Nm\over 2\pi T}\,
      \sqrt {N\over 2} \\
  &\qquad\qquad\qquad\qquad\qquad\qquad \times 
   \exp{(-i) \,{mN^2\over 2T} 
        \left(1 - \cos {p\pi\over N} - {\omega^2 T^2\over 2N^2}\right) q_p^2} 
\end{align*}
The exponent here has sign opposite to that of the usual harmonic
oscillator.  It bears repeating that this does not cause any problems
in the real time formalism used here.\footnote{
In fact, as discussed in section \ref{general}, the integral could
even have been calculated as a distribution in the Euclidean formalism despite the 
exponentially unbounded behaviour of the integrand in that case.}  
The general definition in section \ref{general} gives a 
well-defined meaning to the path 
integral as a  distribution.  
More concretely, the above integrals have an unambiguous 
meaning as the Fourier transforms of distributions, and can 
be operationally evaluated in a non-perturbative 
setting by approximating the integrand by a
convenient family of test functions.  More precisely, the Fourier 
transform of the integrand gives a distribution that is locally equivalent to a 
unique continuous function in a neighbourhood of zero.  The 
above integral is then by definition the
value of this function at zero.\footnote{The integral in each
variable
can easily be checked to exist as an improper Riemann integral,
though it does not have meaning as a Lebesgue integral.  
However, in the present case where the integral is multi-dimensional, an
iterated Riemann
integration would not be invariant under general 
changes of variables \cite{dewitt}.  This, and the 
fact that in general the integrand in the real-time formalism
is a proper distribution, forces us to define the integral in terms of
a transform of a distribution. } 

According to the discussion of the section \ref{general}, the answer,
here calculated for real $T$, is expected to be analytically extensible
to a function on the lower half plane
that may have singularities and cuts on the real axis.  
The path integral expression will give the exact result
as long as we keep careful track of overall factors of $i$.  

To this end, note that for each $p$ such that
$$
  1 - \cos {p\pi\over N} - {\omega^2 T^2\over 2N^2} > 0,
$$
the corresponding integral gives a factor
$$
 1 \left/ {\sqrt{2\left|1 - \cos {p\pi\over N}
       -{\omega^2 T^2\over N^2} \right|}}\right. ,
$$
the absolute value being superfluous here, whereas for 
\begin{align}
  1 - \cos {p\pi\over N} - {\omega^2 T^2\over 2N^2} < 0,
 \label{neg}
\end{align}
we obtain a factor 
$$
 i \left/ {\sqrt{2\left|1 - \cos {p\pi\over N}
       -{\omega^2 T^2\over N^2} \right|}}\right. .
$$
One can check that (\ref{neg}) is satisfied for no values 
of $p$ as $N \to \infty$ when 
$$
  0 < T < {\pi / \omega},
$$
for the single value $p = 1$ when
$$
   {\pi / \omega} < T < {2\pi / \omega},
$$
for the two values $p=1$ and $p = 2$ when 
$$
  {2\pi / \omega} < T < {3\pi / \omega},
$$
and so on.  
We obtain
\begin{align*}
   &i^{\lfloor {T\omega/\pi}\rfloor} \,e^{i\pi/ 4} \lim_{N\to\infty} \sqrt{Nm\over 2\pi T}\,
     \left/\sqrt{\left|\prod_{p=1}^{N-1} {{2\left(1 - \cos {p\pi\over N}
       -{\omega^2 T^2\over N^2} \right)}}
\right|} \right.  \\
&\quad = i^{\lfloor {T\omega/\pi}\rfloor} \,
   e^{i\pi/ 4} \sqrt{m\over 2\pi T} \,\times \\ 
  &\quad\quad\quad \times \lim_{N\to\infty} \left\{\sqrt{N}\left/
     \prod_{p=1}^{N-1} {\sqrt{2\left(1 - \cos {p\pi\over N}
      \right)}}\right. \right\} \left/
     { 
       \sqrt{\left|\prod_{p=1}^{N-1}{ \left(1 - {\omega^2 T^2 \over
       2N^2 \left(1 - \cos {p\pi\over N}\right)}\right)}\right|}}\right. .
\end{align*}
Using the identity
$$
  \prod_{p=1}^{N-1} \left(1 - \cos {p\pi\over N}\right) = {N\over 2^{N-1}},
$$
the factor in braces is seen to be unity.  Furthermore, it can be 
shown that, as $N\to \infty$, one gets
\begin{align*}
  \prod_{p=1}^{N-1}\left(1 - {\omega^2 T^2 \over
       2N^2 \left(1 - \cos {p\pi\over N}\right)}\right) 
   &\to \prod_{p=1}^{N-1}\left(1 - {\omega^2 T^2\over p^2 \pi^2}\right) \\
     &= {\sin \omega T \over\omega T}, 
\end{align*}
and we obtain the final result
\begin{align}
   \braket{x=0, T}{x=0, 0} = i^{\lfloor {T\omega/\pi}\rfloor} \,e^{i\pi/4}\sqrt{m\omega\over 2\pi|\sin \omega T|}.
\label{harmonicvac}
\end{align}
This is the physical result, valid for purely real $T$.  
We can express it as the 
boundary value when approaching the real axis from above
of a function analytic on the whole complex plane with cuts on the 
real axis as
\begin{align}
 \braket{x=0, T}{x=0, 0} = e^{i\pi/ 4} \sqrt{m\omega\over 2\pi\sin \omega\, (T + i0)},   \label{totalvacvac}
\end{align}
Here the meaning of the square root is defined by 
specifying the cuts to be on the intervals 
$$
  (2n-1)\, \pi/\omega < T < 2n\pi/\omega, \quad n \in \mathbf{Z}.  
$$  
The endpoints of the cuts are precisely integer multiples of the
half-period.  At these times the original position state is 
refocused and again becomes a delta distribution, so that 
singularities are indeed expected according to the general analysis 
of section \ref{general}.   

To see that the overall powers of $i$ are correctly reproduced, note
that 
as one travels horizontally from $x + iy$ to $x + 2\pi + iy$ for 
positive $y$, the 
value of $\sin \left(x + iy\right) = \sin x \cosh y + i \cos x \sinh y$
describes a clockwise half circle around the origin in $\mathbf{C}$,
so that $\sqrt{ \sin \left(x + iy\right)}$ picks up a factor $-i$, and
$1/ \sqrt{ \sin \left(x + iy\right)}$ picks up a factor $i$, as required.

This has the spectral
representation
\begin{align}
  e^{i\pi/4}\sqrt{m\omega\over 2\pi\sin \omega\, (T+ i0)}
   = \sqrt{m\omega\over \pi} \left(e^{i\left({\omega\over 2}\right)(T + i0)} 
       + \half\, e^{i\left({\omega\over 2} + 2 \omega\right)\, (T + i0)} + \cdots\right), \label{definiteharmonic}
\end{align}
which has the operator interpretation
\begin{align*}
   \braket{x=0, T}{x=0, 0} &= \bra{x=0} e^{-iHT} \ket{x = 0} \\
    &= \sum_{n = 0}^\infty \bra{x=0}  e^{-iHT} \ket{n}\braket{n}{x = 0} \\
&= \sum_{n = 0}^\infty e^{-iE_n T} \,\left|\braket{x = 0}{n}\right|^2,
\end{align*}
in terms of the states (\ref{posstates}) with positive-definite
inner product $\braket{\cdots}{\cdots}$ and negative energies
$E_n = - {\omega\over 2} - n\omega$.  Notice that only the states 
for even $n$ have nonzero overlap with $\ket{x = 0}$, and that the
path integral gives exactly the correct factors
 $|\braket{x = 0}{n}|^2$.  
Also note that the powers of $i$ required for 
correspondence with (\ref{harmonicvac}) 
are correctly reproduced due to  the common factor 
$$
  e^{i{\omega T / 2}}.
$$
The series converges in the sense of 
Abel summation, which is equivalent to having the $+i0$  
in (\ref{definiteharmonic}).

The positive-inner-product, negative energy spectral
 representation (\ref{definiteharmonic}) is analytically extensible to 
the upper half plane.  As expected from our general discussion in 
section \ref{general}, this 
function can be analytically extended, crossing the real axis
on an interval to the right of the origin, 
to the lower half plane,  where
it can be used to obtain the indefinite inner product, positive energy 
representation.  This function has cuts on the real axis and is given by 
$$
  e^{i\pi/4}\sqrt{m\omega\over 2\pi\sin \omega\, z}
$$
with the constraint that the cut starting at $z = 0$ is taken to the 
left of the origin.   

The values of the indefinite inner product representation inner product 
for real $T$ are
given by approaching the real axis from below, where we get the 
spectral representation
\begin{align}
 & e^{i\pi/4}\sqrt{m\omega\over 2\pi\sin \omega \,(T - i0)} \nonumber\\
   &\quad = i\,\sqrt{m\omega\over \pi} \left(e^{-i\left({\omega\over 2}\right) \, (T - i0)} 
       + \half\, e^{-i\left({\omega\over 2} + 2 \omega\right) \,(T- i0)} + \cdots\right), \label{indefharmonic} 
\end{align}
for all real $T$.    
This has the operator interpretation
\begin{align}
   i\bbraket{x=0, T}{x=0, 0} &= i\bbra{x=0} e^{-iHT} \kket{x = 0}\nonumber \\
&=i \sum_{n = 0}^\infty \bbra{x=0}  e^{-iHT} \kket{n}\bbraket{n}{n}\bbraket{ n}{x = 0}\nonumber \\
&=i \sum_{n = 0}^\infty e^{-iE_nT} \,\bbraket{n}{n}|\bbraket{x = 0}{n}|^2, \label{opint}
\end{align}
in terms of the states   (\ref{negstates})
 with indefinite
inner product $\bbraket{\cdots}{\cdots}$ and positive energies
$E_n = {\omega\over 2} + n\omega$.  Again, the $-i0$ in
(\ref{indefharmonic}) is equivalent to Abel summation. 
We see that the path integral has exactly supplied the 
correct prefactor $i$ to satisfy the relation 
$$
  \braket{x=0, T}{x=0, 0} =  i\bbraket{x=0, T}{x=0, 0}, 
$$
described in our general analysis of section \ref{general}.

Note that  the
path integral gives the correct factors
\begin{align*}
  |\bbraket{x=0}{ 0}|^2 &= \sqrt{m\omega\over\pi},
\end{align*} and 
$$
  \bbraket{n}{n} = +1, \quad \text{$n$ even},
$$
since only the states 
for even $n$ have nonzero overlap with $\bbra{x = 0}$.

We now  calculate the two-point function in the path integral 
approach.  We will first do the calculation for finite
time $T$ and fixed boundary conditions $x(0) = x(T) = 0$,
and then show that the correct ground state expectation value is 
unambiguously extracted from the result.

Consider the two-point function first in momentum space 
\begin{align*}
   e^{i\pi/ 4} \lim_{N\to\infty} \sqrt{Nm\over 2\pi T}\,
     &\biggl\{\mathop{\prod_{p=1}^{N-1}}_{p\ne \tilde p} \int dq_p\, e^{i\pi/ 4}\,\sqrt{Nm\over 2\pi T}\,
      \sqrt {N\over 2} \\
  &\qquad \times 
   \exp{(-i) \,{mN^2\over 2T}
        \left(1 - \cos {p\pi\over N} - {\omega^2 T^2\over 2N^2}\right) q_p^2}
\biggr\} \\
&\times
\int dq_{\tilde p}\, e^{i\pi/ 4}\,\sqrt{Nm\over 2\pi T}\,
      \sqrt {N\over 2} \\
&\qquad \times \, q_{\tilde p}^2\,\,
  \exp{(-i) \,{mN^2\over 2T}
        \left(1 - \cos {p\pi\over N} - {\omega^2 T^2\over 2N^2}\right) q_p^2}.
\end{align*}
The integrands, including the expression on the last line,
are distributions.  Again, the integrals have a rigorous and  
unambiguous interpretation in terms of Fourier transforms of 
 distributions.  Evaluating these, we obtain the momentum space
two-point function
\begin{align*}
  &- e^{i\pi/4}\, \sqrt{m\omega\over 2\pi\, \sin\omega \, (T+i0)}\,
     \lim_{N\to\infty}\left(iT\over mN^2\right)\left/
   \left(1 - \cos {\tilde p \pi\over N} - {\omega^2 T^2\over 2N^2}\right)
   \right. \\
  &\quad = - e^{i\pi/4}\, \sqrt{m\omega\over 2\pi\, \sin\omega \, (T+i0)}\,
     \left(2i\over mT\right){1\over
   (\tilde p \pi/T)^2- \omega^2}. 
\end{align*}
Transforming to position space using (\ref{change}), we obtain
\begin{align}
&\bra{x=0, T} T\, x(t_2)\, x(t_1) \ket{x=0, 0}\nonumber \\
&\quad = 
  - e^{i\pi/4}\, \sqrt{m\omega\over 2\pi\, \sin\omega \, (T+i0)}\,
     \left(2i\over mT\right)\sum_{p=1}^\infty{\sin {p\pi t_1\over T}\, \sin {p\pi t_2\over T}\over
   (\tilde p \pi/T)^2- \omega^2} \nonumber \\
 &\quad = 
  - e^{i\pi/4}\, \sqrt{m\omega\over 2\pi\, \sin\omega \, (T+i0)}\,
     \left(2i\over mT\right)\sum_{p=1}^\infty{\cos {p\pi (t_1 - t_2)\over T} -  \cos {p\pi (t_1 + t_2 - T)\over T}\over
   (\tilde p \pi/T)^2- \omega^2} \nonumber \\
   &=   - e^{i\pi/4}\, \sqrt{m\omega\over 2\pi\, \sin\omega \, (T+i0)}\,
       {i\over 2m\omega} \, \left({-\cos \omega \,(T - |t_2 - t_1|) + \cos \omega\,
(t_2 + t_1)\over \sin \omega T}\right), \label{fixedtwopoint}
\end{align}
where we have used \cite{BG}, formula 1.445 (7) to evaluate the sums 
explicitly.  The sum converges as long as $\omega T\over \pi$ is not
an integer.  At these values we obtain poles, corresponding to the 
refocusing of the initial delta function by the harmonic
oscillator potential after an 
integral number of half-periods, which makes the overlap with the final
state diverge.  

It is important to note that the above summations would not converge 
for imaginary $T$.  The path integral result must
 be fully evaluated for real
$T$ \textit{before} taking the analytic continuation in $T$ to extract the 
ground state expectation values below.

We have obtained the two-point function for real
time and fixed boundary conditions in the positive-definite inner product 
representation, which has no ground state.  
As discussed
 in section \ref{ground}, we may therefore extract the ground state 
expectation value appropriate to the \textit{indefinite} inner product 
representation from the analytic continuation to the lower half plane
in $T$ as
\begin{align}
\bbra{0}  T\, x(t_2)\, x(t_1) \kket{0} &=
   \lim_{T\to-i\infty} {\bra{x=0, T} T\, x(t_2)\, x(t_1) \ket{x=0, -T} \over
    \braket{x=0, T}{x=0, -T}}\nonumber \\
    & = - {1\over 2m\omega}\, e^{-i\omega |t_2-t_1|},  \label{twopoint2}
\end{align}
which agrees with the operator formalism result
(\ref{twopoint}) in the indefinite 
inner product representation.  

From the above and  
$$
  b = \sqrt {m\omega\over 2} \left(x + {i\dot x\over \omega}\right),
\quad  b^\dagger = \sqrt {m\omega\over 2} \left(x - {i\dot x\over \omega}\right),
$$
we easily obtain the desired
$$
 \bbra {0} T \,{b}(t)\, b^\dagger(0) \kket{0}  =
- \theta(t)\, e^{-i\omega t}
  = {1\over i} \int {dk\over 2\pi}\,{e^{-ikt}\over k - \omega + i0}.
$$

\section{The pole prescription is not derivable from a convergence 
factor}
\label{polesect}

It is important to note that the 
two-point function 
(\ref{twopoint2}), which we have unambiguously calculated starting from the 
real-time path integral, does \textit{not} coincide with the naive result one
would obtain if one were to replace
$$
  \omega \to \omega + i\epsilon
$$
in  the path integral 
expression (\ref{fullPI}), as is sometimes incorrectly argued to be
necessary to make the integral converge.  In fact, such a replacement is not 
necessary since, as we have pointed out, the integral already 
exists in the distributional sense without such an $i\epsilon$ factor.  
Instead, the correct and unambiguous result (\ref{twopoint2}) so
obtained for the 
two-point function can be written as
$$
  - {1\over 2m\omega}\, e^{-i\omega |t_2-t_1|}
  = - {1\over im} \int{dk\over 2\pi}\, {e^{-ikt}\over -k^2 + \omega^2 - i0},
$$
which is what one would obtain by inverting the quadratic part of an 
action with the 
\textit{opposite}
replacement
$$
  \omega\to \omega-i0.
$$
Although little practical advantage would have been gained in this 
example by doing so,
we could in fact have reproduced the correct two-point function by adding a
term $-i\epsilon x^2$
to the original action (\ref{fullPI}).   
Adding such a term would make the originally bounded integrand 
exponentially unbounded of quadratic order, so that
this is not a convergence factor.
As explained in section \ref{general}, despite its exponential growth,
such a path integral 
is calculable as a Gel'fand-Shilov distribution, and 
it will indeed give the 
correct result as $\epsilon\to 0$.  However, the fact that this is 
not a ``convergence factor'' shows that the usual naive convergence 
argument will not work here.  

This observation is closely related to the fact that the ordinary 
Euclidean rotation of the path integral (\ref{fullPI}) produces
an integrand of quadratic exponential \textit{growth}.  In other words,
in contrast to the usual case,
the Euclidean path integral is worse than the real-time path integral.  
Again, despite this bad behaviour, such a Euclidean path integral
can be calculated as a distribution
using the methods of section \ref{general}, and the correct results 
will be obtained.

\section{Combining harmonic and phantom harmonic oscillators}
\label{combining}

In this section we briefly describe the two-particle system consisting of 
a harmonic oscillator and a phantom harmonic oscillator, with 
Hamiltonian
$$
  H = {p_1^2\over 2m} + {m\omega^2\over 2}\,x_1^2
      - {p_2^2\over 2m} - {m\omega^2\over 2}\,x_2^2.
$$
This system is instructive in exhibiting some features not seen in the
examples we have considered so far.  In particular, in the 
positive definite Hilbert space quantization
\begin{itemize}
\item
  The spectrum 
  is neither bounded above nor bounded below.
\item
  The transition amplitude exists for all real $T$ except for 
  simple pole singularities (not cuts), but
  is not the boundary value of any analytic function.  
\item
  Even so, the restriction of the amplitude to a suitable initial 
interval in $T$ can be extended to an analytic function on the lower 
half plane, and indeed to the entire complex plane except for the
  poles on the real axis.
\end{itemize}
In the indefinite representation, on the other hand
\begin{itemize}
\item
  The spectrum has a ground state.
\item
  The transition amplitude 
  is the boundary value of an analytic function extensible to the lower
  half plane, and indeed to the entire complex plane except for 
  poles on the real axis.
\end{itemize}
Still the arguments of section \ref{general} will hold.  The transition 
amplitude for the two representations agree for initial real $T$, so that
ground state expectation values appropriate to the indefinite representation
can be extracted from either quantization. 

To obtain the transition amplitudes, we use the following known result for the
ordinary harmonic oscillator
$$
\braket{x_1=0, T}{x_1=0, 0} = 
  e^{-i\pi/4}\sqrt{m\omega\over 2\pi\sin \omega\, (T- i0)},
$$
which, combined with the result 
(\ref{definiteharmonic})
$$
\braket{x_2=0, T}{x_2=0, 0} = 
  e^{i\pi/4}\sqrt{m\omega\over 2\pi\sin \omega\, (T+ i0)},
$$
for the phantom harmonic oscillator in the positive definite representation, gives
$$
\braket{x_1=0, x_2=0, T}{x_1=0, x_2=0, 0} =
(-)^{\lfloor {T\omega/\pi}\rfloor}\,
  {m\omega\over 2\pi\sin \omega T}.
$$
We emphasize that, even though the system does not have a ground state, the 
real-time transition amplitude is well-defined.  However, due to the 
factor $(-)^{\lfloor {T\omega/\pi}\rfloor}$, we do not obtain the boundary value of 
any analytic function -- the function cannot be analytically extended to 
either the upper or lower half plane in $T$, which is as expected in the absence
of either bound on the energy.  We also point out that 
the singularities on the real line are not cuts now,
but poles.  

On the other hand, choosing the indefinite representation for the phantom 
oscillator, while keeping the definite representation for the ordinary oscillator,
 the combined system has a ground state.  Since
$$
i\bbraket{x_2=0, T}{x_2=0, 0} = 
  e^{i\pi/4}\sqrt{m\omega\over 2\pi\sin \omega\, (T- i0)},
$$
we find
$$
i\bbraket{x_1=0, x_2=0, T}{x_1=0, x_2=0, 0} = 
  {m\omega\over 2\pi\sin \omega T},
$$ 
where 
$$\bbraket{\phi, \psi}{\phi', \psi'} \equiv
      \braket {\phi}\phi'\,\bbraket {\psi}\psi'.$$
The transition function can be analytically extended to the lower half plane, as expected
in the presence of a lower bound on the energy. 

Consistent with the arguments of section \ref{general}, the two results
agree for an initial real interval, concretely $0<T<\pi/\omega$, which allows us to obtain the result
for the indefinite representation from the calculation in the definite representation.

\section{The Dirac boson}
\label{dirac}

In this section we discuss the quantization of a Dirac field
with bosonic statistics.  This kind of field arises, for example, 
as regulator fields in 
Pauli-Villars and Lee-Wick regularizations of theories 
containing Dirac fermions.

This field is particularly interesting, since its 
Euclidean functional integrand is exponentially unbounded.
Despite this, both the real-time and the Euclidean functional integrals
may be computed using the methods of section \ref{general}.  We will 
do the computation in the real-time formalism.  

Relativistic field theory requires locality (causality) and existence of 
a ground state.
These conditions, imposed on a bosonic Dirac field, requires 
that we choose an indefinite inner product representation for anti-particle
modes.  However, we shall see that 
the indefinite representation is unsuitable for 
formulating a covariant
functional integral, since there is no 
covariant way of separating these modes when we are integrating 
over all configurations, including those that are not 
solutions to the equations of motion.  

For this reason, it will
be very useful to us that, as we have argued,
 we can extract the same results from the 
manifestly covariant functional integral based on the definite
representation.  
 
Our starting point is the Dirac action
\begin{align}
  \mathcal{S} = \int \bar \phi \left(i \gamma^\mu\partial_\mu - m\right)\phi,
\label{diracaction}
\end{align}
where we take the field $\phi$ to be bosonic.  We first note that,
because the spinor field is bosonic,
any Euclidean version of this action \cite{waldron} 
is unbounded below, so that 
the Euclidean path integrand is exponentially unbounded.   
Since the field appears quadratically, such an unbounded integrand could 
be treated using the methods of  section \ref{general}.  However, it 
will be easier to work in the real-time formalism.

We may expand
the action in modes as 
\begin{align}
  \mathcal{S} = \int dt\sum_s\int {d^3p\over (2\pi)^3}
\left(i\left({a^s_p}^\dagger \dot a^s_p + {b^s_p}^\dagger \dot b^s_p\right)
 - E_p \left({a^s_p}^\dagger a^s_p - {b^s_p}^\dagger  b^s_p\right)\right), \label{modeaction}
\end{align}
where $a_p^s$ and $b_p^s$ are bosonic.  This leads to the
commutation relations
\begin{align}
  [a^s_p, {a^s_q}^\dagger] = (2\pi)^3 \, \delta^3(p - q)\,\delta^{rs},
  \quad 
[b^s_p, {b^s_q}^\dagger] = (2\pi)^3 \, \delta^3(p - q)\,\delta^{rs}.
\label{fieldcomm}
\end{align}
with respect to which the field satisfies locality \cite{PS}.  
However, if we define the vacuum state by
$$
  a^s_p \ket{0} = b^s_p \ket{0} = 0,
$$
obtaining a positive-definite state space, 
it is easy to see that we may create states with arbitrary negative 
energy by applying suitable anti-particle creation 
operators ${b^s_p}^\dagger$.  

We can obtain a local (causal),  
positive-energy representation  by keeping the 
commutation relations of the modes the same but changing 
the ground state to satisfy
\begin{align}
  a^s_p \kket{0} = {b^s_p}^\dagger \kket{0} = 0.  \label{newgs} 
\end{align}
Since this does not affect the
 commutation relations of the fields $\phi$ and 
$\phi^\dagger$, the desired locality is preserved.  
However, as in section \ref{pharmonic}, the representation
now has an indefinite inner product.  This is a general
feature of local fields with the ``wrong'' statistics,\footnote{But note 
that Pauli-Villars regulator fields are allowed to have the ``wrong''
statistics.}    as 
discussed, for example, in \cite{bogolubov}.
This choice of ground state 
 is equivalent to interchanging what we mean by anti-particle 
creation/annihilation, and we could, if we wished, reflect that
in the notation by calling $b^s_p \leftrightarrow {b^s_p}^\dagger$
everywhere.  

Nothing prevents us from
 constructing a functional integral based on the indefinite 
representation
(\ref{fieldcomm}) and (\ref{newgs}).  However, this cannot be done 
covariantly.  To understand this, note that in the indefinite  
representation
(\ref{newgs}), each $a^s_p$ mode generates a positive-definite
factor in the state space, while each $b^s_p$ mode generates an
indefinite factor.  As a result, 
the two sets of modes would 
have be represented differently in the functional integral
constructed from this representation. 
Specifically, as we saw in section \ref{freesect}, the configuration space 
of a ghost degree of freedom is the imaginary axis, so the 
range of integration for the real and imaginary parts of $b^s_p$ 
would become the imaginary axis \cite{BG, arisue, sakoda},
whereas this would not happen for $a^s_p$. 
However, the mode expansion (\ref{modeaction}) depends on 
a particular space-time decomposition and is not Lorentz-invariant.  
While the particle-antiparticle distinction
can be made covariantly on the space of solutions, a covariant distinction
is not possible in the space of arbitrary time-dependent configurations
of the field used in the functional integral.  
The path integral resulting from the indefinite representation, 
while correct, would not be 
manifestly covariant.

Since the definite representation treats all modes
the same as far as their range of integration is concerned,
 the definite representation can be used to 
generate an ordinary functional integral, based on 
the usual sum over configurations using the Dirac action
(\ref{diracaction}), that does not suffer 
from this lack of covariance.  
Fortunately, we have learned how to relate indefinite and 
definite representations.  

To calculate, note that, due to the sign of the 
energy term, the action for each $b^s_p$ mode in 
(\ref{modeaction}) is equivalent to that of the phantom 
harmonic oscillator (\ref{phantomfo}).  As we did in section
\ref{pharmonic}, we may then extract
the ground state expectation values for the physically
relevant indefinite representation
from the functional integral performed in the 
definite representation.  We find\footnote{Here $b^s_p$ corresponds
to $a$ in (\ref{phantomfo}), which corresponds to $b^\dagger$ in the notation of (\ref{twopoint0}).}
\begin{align*}
  \bbra {0} T \,{b^s_p}^\dagger(t)\, b_p^s(0) \kket{0} 
&= \lim_{T\to -i\infty}
  { \bra {\phi=0,T} T \,{b^s_p}^\dagger(t)\, b_p^s(0) \ket{\phi=0, -T} 
   \over 
     \braket{\phi=0, T}{\phi=0, -T}} \\
&= - \theta(t)\, e^{-iE_p t},
\end{align*}
where the calculation on the right was done
in section \ref{pharmonic} in the 
definite representation. From this, we obtain the correct 
Dirac two-point function, appropriate for the indefinite inner product
theory, as
\begin{align*}
  \bbra{0} T\,\phi(x)\, \bar\phi(y) \kket{0} 
  &= \lim_{T\to -i\infty}
     { \bra{\phi=0, T} T\,\phi(x)\, \bar\phi(y) \ket{\phi=0, -T} 
   \over  \braket{\phi=0, T}{\phi=0, -T}} 
\\
  &= i \int{d^4p\over (2\pi)^4}\, {\gamma^\mu p_\mu + m
        \over p^2 - m^2 + i0} \, e^{-ip\cdot (x - y)},
\end{align*}
where the quantities on the right hand side are 
calculated using the conventional, manifestly covariant, path integral
constructed from the definite representation.\footnote{We should 
qualify this by noting  that while the action used in the functional integrand
is covariant, the boundary conditions are not.}

This two-point function coincides with that 
of a fermionic Dirac particle.  As required, the field satisfies 
locality.    

We note again that we could have reproduced 
this by replacing $m\to m- i\epsilon$ in the
original action (\ref{diracaction}), obtaining an exponentially
unbounded integrand that
could still be calculated in the distributional sense according to the 
discussion of section \ref{general}.  Note, however, that  
the correct replacement $m\to m- i\epsilon$ is opposite to the
naive convergence factor, and that our above real-time calculation 
showed such a replacement to be unnecessary.

\section{Complex energies and non-perturbative 
Lee-Wick type pole prescriptions}
\label{leewicksect}

We consider a simple example where an interaction gives rise 
to complex energies in the indefinite representation.  For this
example, we  
show that the path integral exists and provides a unique 
pole prescription for the relevant contour integrals.

The example consists of a harmonic oscillator $a$ and a phantom 
 harmonic oscillator $b$  satisfying
$$
   [a, a^\dagger] = 1, \quad [b, b^\dagger] = -1,
$$
with hermitian Hamiltonian
$$
  H = \omega\,(a^\dagger a - b^\dagger b) + \gamma\, (a^\dagger b + b^\dagger a).
$$
in an indefinite inner product representation based on a state
$\kket{0}$, where
$$
  a\kket{0} = b \kket{0} = 0, \quad \bbraket{0}{0} = 1.
$$
This system may be solved in the operator formalism by defining
operators
\begin{align*}
 & \bar a^\dagger = a^\dagger + i b^\dagger, \quad \bar a = a - i b \\
 & \bar b^\dagger = a^\dagger - i b^\dagger, \quad \bar b = a + i b, 
\end{align*}
that have commutation relations
\begin{align*}
  &[\bar a, \bar a^\dagger] = 0, \quad  [\bar b, \bar b^\dagger] = 0 \\
&[\bar a, \bar b^\dagger] = 2, \quad  [\bar b, \bar a^\dagger] = 2.
\end{align*}
Here $\bar a^\dagger$ and $\bar b^\dagger$ create null states from 
the vacuum $\kket{0}$.  The Hamiltonian becomes
$$
  H = \half\,(\omega - i\gamma) \,\bar a^\dagger \bar b + 
   \half\,(\omega + i \gamma)\, \bar b^\dagger
    \bar a,
$$
so that the energy eigenvalues are
$$
  E_{m, n} = (\omega - i\gamma)\,m + (\omega + i\gamma)\, n, \quad 
    m, n = 0, 1 , 2 , \dots
$$
The spectrum includes complex eigenvalues.  Since the Hamiltonian is 
hermitian, these come in complex conjugate pairs, as expected in an
indefinite inner product space, and the corresponding eigenstates 
are null.  

We now apply the considerations of section \ref{general} to 
construct the path 
integral representation of this system. 
Since all terms in the action are at most quadratic in the position and 
momentum of either oscillator, and since no terms of the form 
$x_1 p_1$ or $x_2 p_2$ appear, we may obtain the ground state 
expectation values appropriate to the indefinite
inner product representation from a  
path integral calculated as a distribution either on the test space
$$
  S_{1/2}^{1/2}(x_1)\otimes  S_{1/2}^{1/2}(ix_2),
$$
corresponding to the indefinite representation used above, or on 
the test space
$$
  S_{1/2}^{1/2}(x_1)\otimes  S_{1/2}^{1/2}(x_2)
$$
corresponding to a positive-definite inner product representation for the 
$b$-oscillator.  It was the content of section \ref{ground}
that both representations encode the same ground state 
expectation values.   

As we previously discussed, the definite representation
gives a simpler path integral.  
With this choice,
the action used in the path integral 
$$
  S = \half \int dt \left\{\bar a^* \left(i\partial_t - \omega + i \gamma\right) {\bar b} + \bar b^* \left(i\partial_t - \omega - i \gamma\right){\bar a}
   \right\}.
$$
is real-valued, since the path integration measure is the usual
one.

On the other hand, in the indefinite representation, we saw
that the configuration space for each hermitian phantom degree of freedom 
is the imaginary axis.  As a result, the integration 
path in the path integral measure for each of
$b + b^*$ and $i \, (b - b^*)$ is rotated to the imaginary 
axis, causing the above action to be complex-valued.  The 
integrand is therefore exponentially unbounded, in this case of linear order
of growth.  

However, despite this, it was the content of section \ref{general}
that the the corresponding 
path integrals still 
exist as distributions on the 
chosen test spaces.   

Because they lacked a non-perturbative definition of the indefinite
inner product
theories corresponding to such unbounded path integrands, 
various authors \cite{leewick, lee, cutkosky} studied the consistency of 
ad hoc prescriptions for defining
the two-point functions, and more complex diagrams, order by order 
in perturbation theory.  
Lee and Wick \cite{leewick, lee} introduced one such prescription,
 requiring the contour 
integrations defining various diagrams to be continuously deformed 
to avoid the movement of poles in the complex plane
as we increase parameters such as $\gamma$
starting from zero.  Cutkosky, Landshoff, Olive and Polkinghorne \cite{cutkosky}
attempted to generalize this prescription to coalescing 
singularities not covered by Lee and Wick.  Boulware and Gross \cite{BG}
discussed the problem from the path integral point of view but did not 
succeed in defining these theories non-perturbatively, mainly 
due to the unboundedness of the path integrand.  

The framework of this article 
allows a non-perturbative functional integral to be calculated
in the distributional sense  
for various kinds of unbounded path integrands, including the ones
studied by Lee and Wick.  In principle, we expect the ad hoc 
prescriptions of the above authors to be either reproduced or corrected
in our non-perturbative framework.  

We start with the simplest such calculation, where we derive a
prescription of Lee-Wick type for the two-point function from the 
 path integral.  

Consider the problem of obtaining the two-point function
$$
  \eexpect{T \,\bar a(t)\, \bar b^\dagger (0)}
$$
from the above action in the path-integral approach.  
If we were to naively invert the 
relevant quadratic term, we would obtain 
\begin{align*}
  -{2\over i}\int {dk\over 2\pi} \, {e^{-ikt}\over k - (\omega + i\gamma)} 
= -2\, \theta(-t)\, e^{-i\,(\omega+i\gamma)\, t}.
\end{align*}
A glance at the operator representation confirms that this is wrong.
The correct result is reproduced by
\begin{align}
  \eexpect{T \,\bar a(t)\, \bar b^\dagger (0)} 
  = 2\,\theta(t)\, e^{-i\,(\omega+i\gamma)\, t}
   = -{2\over i}\int_\Gamma {dk\over 2\pi}\, {e^{-ikt}\over k - (\omega + i\gamma)}, \label{correctba}
\end{align}
where the integration contour $\Gamma$ has been displaced from the real line
 to pass above the pole at $1 + i\gamma$.
This prescription, in the spirit of Lee and Wick, seems ad 
hoc from a naive path integral point of view, but can be rigorously 
derived in our formalism.  

We will now show that the correct result 
can be unambiguously obtained, without any ad hoc prescription, 
 from a careful evaluation of the 
path integral.  
We already discussed in section \ref{polesect} 
how the unusual pole prescription in the
free two-point function of $b$ follows unambiguously from the path integration.
The current discussion generalizes this to the case of complex energies.   
As we have seen, we need to evaluate the path integral for 
fixed boundary conditions between real times $T$ and $-T$, and then
take the limit $T\to -i\infty$ of the analytic continuation 
of the result to complex $T$.  
Since we have already done this for the free oscillator 
in section \ref{pharmonic}, we will start from those results 
and evaluate the path integral 
by treating the quadratic interaction perturbatively in $\gamma$
starting from the form
$$
  S = \int dt \left(a^\dagger\, (i\partial_t - \omega)\, a
      - b^\dagger\, (i\partial_t - \omega)\, b + 
     \gamma\, (a^\dagger b + b^\dagger a)
    \right)
$$
of the action.  Nothing is
lost in this approach, since the perturbation is quadratic, so that the series can be
exactly summed, giving the exact result for the full path integral.

Our starting point is the result (\ref{fixedtwopoint}), describing 
the two-point function of the free $b$-field given the fixed boundary conditions.  The
two-point function for the $a$-field is identical except for an opposite 
overall sign.  
Performing perturbation theory with fixed boundary conditions
at real $-T$ and $T$, analytically continuing
the result in $T$ and taking
the limit as $T\to -\infty$ as discussed, we find
\begin{align*}
  \eexpect{T \,{a}(t)\, a^\dagger(0)} 
  &= i\biggl[\int {dk\over 2\pi}\, {e^{-ikt}\over k-\omega + i0}
     - \gamma^2 \int {dk\over 2\pi}\, {e^{-ikt}\over (k-\omega + i0)^3} \\
& \qquad \qquad
 + \gamma^4 \int {dk\over 2\pi}\, {e^{-ikt}\over (k-\omega + i0)^5}
  - \cdots
   \biggr] \\
 &= \theta (t)\, e^{-i\omega t}\, \left( 1 - {\gamma^2 (it)^2 \over 2!}
      + {\gamma^4 (it)^4 \over 4!} - \cdots\right) \\
&= \theta (t)\, e^{-i\omega t} \,\cosh \gamma t \\
&= \half\, \theta (t) \left(e^{-i\,(\omega + i\gamma)\,t}
        + e^{-i\,(\omega - i\gamma)\,t} \right) \\
&= \half \,i \int_\Gamma {dk\over 2\pi}\, 
  \left[{e^{-ikt}\over k-\omega - i\gamma}
       + {e^{-ikt}\over k-\omega + i\gamma}\right],
\end{align*}
where the integral contour $\Gamma$ has been deformed from the real axis
into the upper half plane
to pass above the 
pole at $\omega + i\gamma$ to reproduce the exact path integral result
on the previous line.  There is no need to postulate this contour
prescription ad hoc, since we have obtained it from an exact path integral 
calculation using the 
non-perturbatively defined path integral.\footnote{Note 
that we would have obtained a wrong answer 
if we had interchanged 
the order if integration and summation, first
formally summing the geometric series in momentum space and only 
afterwards taking the Fourier transform.  We are prevented from 
doing this by the fact that the momentum-space geometric series
does not converge for the full range of $k$ over which we integrate.  
The above answer is 
 unambiguous. }  

The expectation values 
$$\eexpect{T \,{b}(t)\, b^\dagger(0)}, \quad\eexpect{T \,{b}(t)\, a^\dagger(0)}, \quad
\eexpect{T \,{a}(t)\, b^\dagger(0)},
$$
 may be similarly 
computed and put together to reproduce the 
exact operator formalism result (\ref{correctba}).

This method generalizes to more complex diagrams in interacting
theories extending the above quadratic action.  
Consider, for example, an $a$-$b$
 loop diagram, expected to be proportional to
$$
  \int_? {dk\over 2\pi} \left({1\over k-\omega + i\gamma} +
       {1\over k-\omega - i\gamma} 
        \right) \left(
    {1\over p - k -\omega - i\gamma} + {1\over p - k -\omega + i\gamma}\right),
$$
where the integration contour is so far unspecified.
As above, we now do an exact, non-perturbative path integral 
calculation that determines the 
appropriate integration contour.  The path integral may be 
unambiguously calculated for real $T$ and the
expectation value extracted by continuing $T\to -i\infty$ is unique. 
Taking this limit term by term in the perturbation 
series around $\lambda = 0$ gives
\begin{align*}
&\int {dk\over 2\pi}\, {1\over k-\omega + i0} \,
    {1\over p - k -\omega + i0}
- \gamma^2 \int {dk\over 2\pi}\, {1\over (k-\omega + i0)^3} \,
    {1\over p - k -\omega + i0} \\
&\qquad
- \gamma^2\int {dk\over 2\pi}\, {1\over k-\omega + i0} \,
    {1\over (p - k -\omega + i0)^3}
+ \cdots,
\end{align*}
where the integrals are over the real line.  Again, we may not 
arbitrarily exchange sums and integrations.  The integrals may 
be calculated by closing the contours in either the upper or lower half
plane.  Choosing the latter, only the poles contributed by the 
left factors contribute, and denoting
$$
  f(k) \equiv {1\over p - k -\omega + i0},
$$
 the result is proportional to
\begin{align*}
 & \left[f(\omega) - \gamma^2 \,f''(\omega) + \gamma^4\, f''''(\omega) 
     - \cdots\right] \\
 &\qquad \qquad - \gamma^2 \left[
   f^3(\omega) - \gamma^2 \,(f^3)''(\omega) + \gamma^4\, (f^3)''''(\omega) 
     - \cdots
    \right] \\
&\qquad \qquad + \gamma^4 \left[
   f^5(\omega) - \gamma^2 \,(f^5)''(\omega) + \gamma^4\, (f^5)''''(\omega) 
     - \cdots
    \right] \\
&\qquad \qquad  + \cdots \\
&\qquad\propto f (\omega + i\gamma) + f (\omega - i \gamma) 
 -\gamma^2 \left[f^3 (\omega + i\gamma) + f^3 (\omega - i \gamma) \right]  
  + \cdots \\
&\qquad= {f(\omega + i\gamma) \over 1 + \gamma^2 \,f^2 (\omega + i\gamma)}
    + {f(\omega - i\gamma) \over 1 + \gamma^2 \,f^2 (\omega - i\gamma)} \\
 &\qquad\propto {2\over p - 2\omega} + {1\over p - 2\,(\omega + i\gamma)}
    + {1\over p - 2\,(\omega - i\gamma)},
\end{align*}
since $|\gamma^2 \,f^2 (\omega \pm i\gamma)| \le 1$ so that the geometric 
series converge.  The poles are precisely at the positions one would expect from 
the various possible combinations of two intermediate particles with energies 
$\omega \pm i\gamma$.  This result is reproduced by the contour integral
$$
  \int_\Gamma {dk\over 2\pi} \left({1\over k-\omega + i\gamma} +
       {1\over k-\omega - i\gamma} 
        \right) \left(
    {1\over p - k -\omega - i\gamma} + {1\over p - k -\omega + i\gamma}\right),
$$
where the integration path $\Gamma$ stretches from $-\infty$ to $\infty$, but is
deformed to pass below the two poles at $p - \omega \pm i\gamma$ and above the 
two poles at $\omega \pm i\gamma$.  This coincides with Lee and Wick's 
prescription that the contour should be deformed so that no poles cross it
as we change $\gamma$ continuously starting from $\gamma =0$.  We have derived
it unambiguously as a non-perturbative result in the path integral 
formulation.

We have seen that Lee-Wick type contour prescriptions can 
be obtained unambiguously from a non-perturbative path integral.
  Although we have only considered 
two very simple toy examples, there is every reason to
expect that similar non-perturbative calculations
 can be used to either reproduce or
correct the 
prescriptions of
Cutkosky, Landshoff, Olive and Polkinghorne unambiguously.

\section{Conclusion}

In this article, we investigated
 the non-perturbative quantization of phantom and ghost 
degrees of freedom by relating their representations in definite
and indefinite inner product spaces.  For a large class of potentials,
we argued that the same physical information can be extracted from 
either representation, and  provided a non-perturbative definition of
the path integral for these theories.   We 
applied the results to the study of ghost fields of Pauli-Villars
and Lee-Wick type, calculating  non-perturbatively 
previously ad hoc prescriptions for Feynman diagram contour integrals 
in the presence  of complex energies. 

The initial motivation for this work was to understand the 
nonperturbative path integral quantization of certain bosonic 
Pauli-Villars ghosts for which the Euclidean path 
integrand is unbounded \cite{myself}.     These ghosts were an important ingredient in the 
construction of a class of 
functional integral measures that are generally
 covariant, background-independent
and conformally invariant.  The current article provides the missing
ingredient to make the construction of \cite{myself} rigorous.

The mathematical framework of this article may have applications
in the study of phantoms appearing  in certain cosmological 
models of dark energy. 
It may also be of interest in the approaches to the cosmological 
constant problem based on phantoms,
or in the approaches based on symmetries consisting of rotations of the 
configuration space to the imaginary axis.   

We pointed out some pitfalls in trying to derive $i\epsilon$
prescriptions in Feynman integrals from naive convergence 
terms in the action.  We showed that for ghost fields, the 
correct prescription is in fact opposite to what one would
obtain from such convergence factors.    

Finally, the methods introduced here may conceivably be useful 
in the complex-coordinate or complex-momentum approaches 
to the study of resonant scattering.

\section*{Acknowledgments}

    We would like to thank Prof.\ Antal Jevicki and the Brown
    University Physics department for their support.

\end{document}